\documentstyle[sprocl,psfig]{article}

\catcode`\@=11
\def\@cite#1#2{{\if@cghi$\!^{#1}$\else$[{#1}]$\fi\if@tempswa\typeout
        {IJCGA warning: optional citation argument
        ignored: `#2'} \fi}}
\catcode`\@=\active

\def\pl#1#2#3{Phys.~Lett.~{\bf {#1}B} (19{#2}) #3}

\def\np#1#2#3{Nucl.~Phys.~{\bf B{#1}} (19{#2}) #3}

\def\cites#1#2{\cite{#1}$^-\,$\cite{#2}}
\def\wt{\widetilde}

\def\sst{\scriptscriptstyle}
\def\ra{\rightarrow}
\def\alg{$shs^E(8|4)$}

\def\del{\partial}

\def\be{\begin{equation}}\def\ee{\end{equation}}
\def\ba{\begin{array}}\def\ea{\end{array}}
\def\bea{\begin{eqnarray}}\def\eea{\end{eqnarray}}
\def\bd{\begin{document}}\def\ed{\end{document}}
\def\fin{\end{thebibliography}\end{document}}

\let\la=\label

\let\bm=\bibitem
\def\nn{\nonumber}
\def\qq{\quad\quad}
\let\fr=\frac

\def\ft#1#2{{\textstyle{{\scriptstyle #1}\over {\scriptstyle #2}}}}
\def\fft#1#2{{#1 \over #2}}
\def\sst#1{{\scriptscriptstyle #1}}
\def\oneone{\rlap 1\mkern4mu{\rm l}}
\newcommand{\eq}[1]{(\ref{#1})}
\newcommand{\w}[1]{\\[0.#1cm]}

\def\Dot#1{\buildrel{_{_{\hskip 0.01in}{\normalsize\bullet}}}\over{#1}}
\def\Hat#1{\widehat{#1}}
\def\bsh{\backslash}
\def\dA{\Dot A}
\def\dB{\Dot B}
\def\dC{\Dot C}
\def\a{\alpha}\def\adt{\dot \alpha}
\def\b{\beta}\def\bdt{\dot \beta}
\def\c{\gamma}\def\C{\Gamma}\def\cdt{\dot\gamma}
\def\d{\delta}\def\D{\Delta}\def\ddt{\dot\delta}
\def\e{\epsilon}\def\vare{\varepsilon}
\def\f{\phi}\def\F{\Phi}\def\vvf{\f}
\def\h{\eta}
\def\k{\kappa}
\def\l{\lambda}\def\L{\Lambda}
\def\m{\mu}
\def\n{\nu}
\def\p{\pi}
\def\P{\Pi}
\def\r{\rho}
\def\s{\sigma}\def\S{\Sigma}
\def\t{\tau}
\def\th{\theta}\def\Th{\Theta}\def\vth{\vartheta}\def\tb{\bar\theta}
\def\X{\Xeta}
\def\x{\xi}
\def\o{\omega}\def\O{\Omega}

\def\ua{\underline{\alpha}}
\def\ub{\underline{\phantom{\alpha}}\!\!\!\beta}
\def\uc{\underline{\phantom{\alpha}}\!\!\!\gamma}
\def\um{\underline{\mu}}
\def\ud{\underline\delta}
\def\ue{\underline\epsilon}
\def\una{\underline a}\def\unA{\underline A}
\def\unb{\underline b}\def\unB{\underline B}
\def\unc{\underline c}\def\unC{\underline C}
\def\und{\underline d}\def\unD{\underline D}
\def\une{\underline e}\def\unE{\underline E}
\def\unf{\underline{\phantom{e}}\!\!\!\! f}\def\unF{\underline F}
\def\ung{\underline g}
\def\unm{\underline m}\def\unM{\underline M}
\def\unn{\underline n}\def\unN{\underline N}
\def\unp{\underline{\phantom{a}}\!\!\! p}\def\unP{\underline P}
\def\unH{\underline{H}}
\def\unF{\underline{F}}\def\unT{\underline{T}}
\def\ovA{\overline{A}}\def\ovB{\overline{B}}
\def\uC{{\underline C}}

\def\smpl{ +}
\def\smm{ -}
\def\ns{\normalsize}
\def\vs{\vspace{-0.25cm}}

\def\bull#1#2#3{\put(#1,#2)
{\makebox(0,.5){\circle*{2}}\makebox(10,0){$\ #3$}}}
\def\ebull#1#2#3{\put(#1,#2)
{\makebox(0,.5){\circle {2}}\makebox(10,0){$\ #3$}}}
\def\dia#1#2#3{\put(#1,#2)
{\makebox(0,0){$\times$}\makebox(13,0){$\ #3$}}}
\def\vgrid#1{\put(#1,0){\line(0,1){2}}}
\def\hgrid#1{\put(-20,#1){\line(1,0){2}}}

\def\fa{
\begin{figure}[!ht]
\unitlength=.7mm
\centerline{
\begin{picture}(100,180)(-20,-10)
\put(-20,0){\line(1,0){150}}
\put(-20,0){\line(0,1){195}}
\put(-25,180){\makebox{${\bf D}$}}
\put(-25,165){\makebox{$11$}}
\put(-25,150){\makebox{$10$}}
\put(-25,135){\makebox{$9$}}
\put(-25,120){\makebox{$8$}}
\put(-25,105){\makebox{$7$}}
\put(-25,90){\makebox{$6$}}
\put(-25,75){\makebox{$5$}}
\put(-25,60){\makebox{$4$}}
\put(-25,45){\makebox{$3$}}
\put(-25,30){\makebox{$2$}}
\put(-25,15){\makebox{$1$}}
\put(120,-5){\makebox{${\bf p}$}}
\put(100,-5){\makebox{$5$}}
\put(80,-5){\makebox{$4$}}
\put(60,-5){\makebox{$3$}}
\put(40,-5){\makebox{$2$}}
\put(20,-5){\makebox{$1$}}
\put(0,-5){\makebox{$0$}}
\bull{0}{135}{}
\bull{20}{150}{}
\bull{40}{165}{}
\bull{0}{75}{}
\bull{20}{90}{}
\bull{40}{105}{}
\bull{60}{120}{}
\bull{80}{135}{}
\bull{100}{150}{}
\bull{0}{45}{}
\bull{20}{60}{}
\bull{40}{75}{}
\bull{60}{90}{}
\bull{0}{30}{}
\bull{20}{45}{}
\bull{40}{60}{}
\vgrid{0}\vgrid{20}\vgrid{40}\vgrid{60}\vgrid{80}\vgrid{100}
\hgrid{15}\hgrid{30}\hgrid{45}\hgrid{60}\hgrid{75}
\hgrid{90}\hgrid{105}\hgrid{120}\hgrid{135}\hgrid{150}
\end{picture}}
\caption{A minimal $p$-brane scan.  These are $p$-branes in $D$ dimensions for
which the collective coordinates form worldvolume scalar supermultiplets.}
\end{figure}
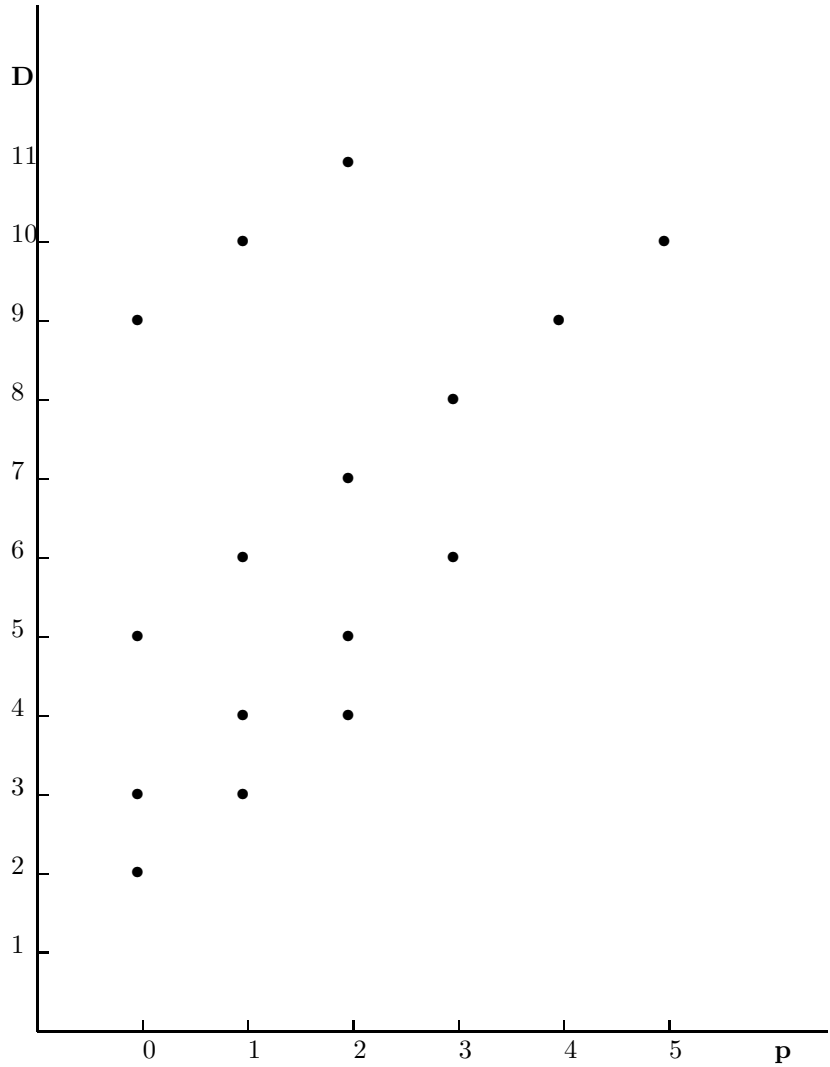}

\def\fb{
\begin{figure}
\centerline{\phantom{xxxxxxxx}\psfig{figure=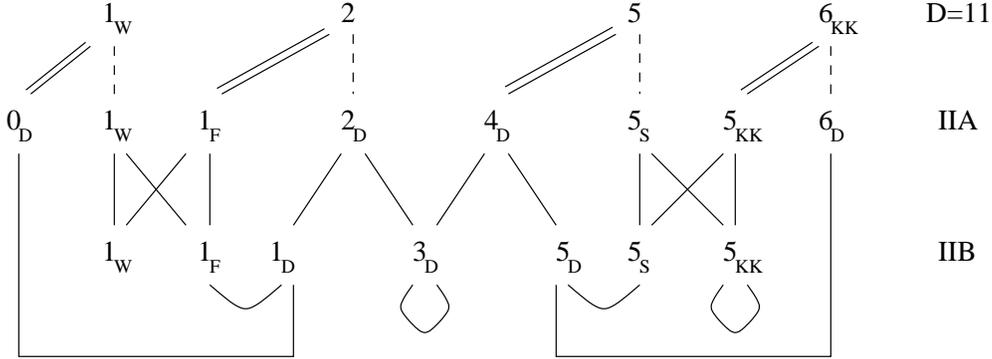}}
\caption{Dimensional reduction and S/T duality maps. The straight
arrows between type IIA/B branes denote T dualities, curved lines
between the type IIB branes denote S dualities, the dashed denote vertical
dimensional reduction and double lines denote double dimensional reduction.
The $Dp$ branes for $p=-1,7,8,9$ are special cases which are not shown.}
\end{figure}}

\def\fc{
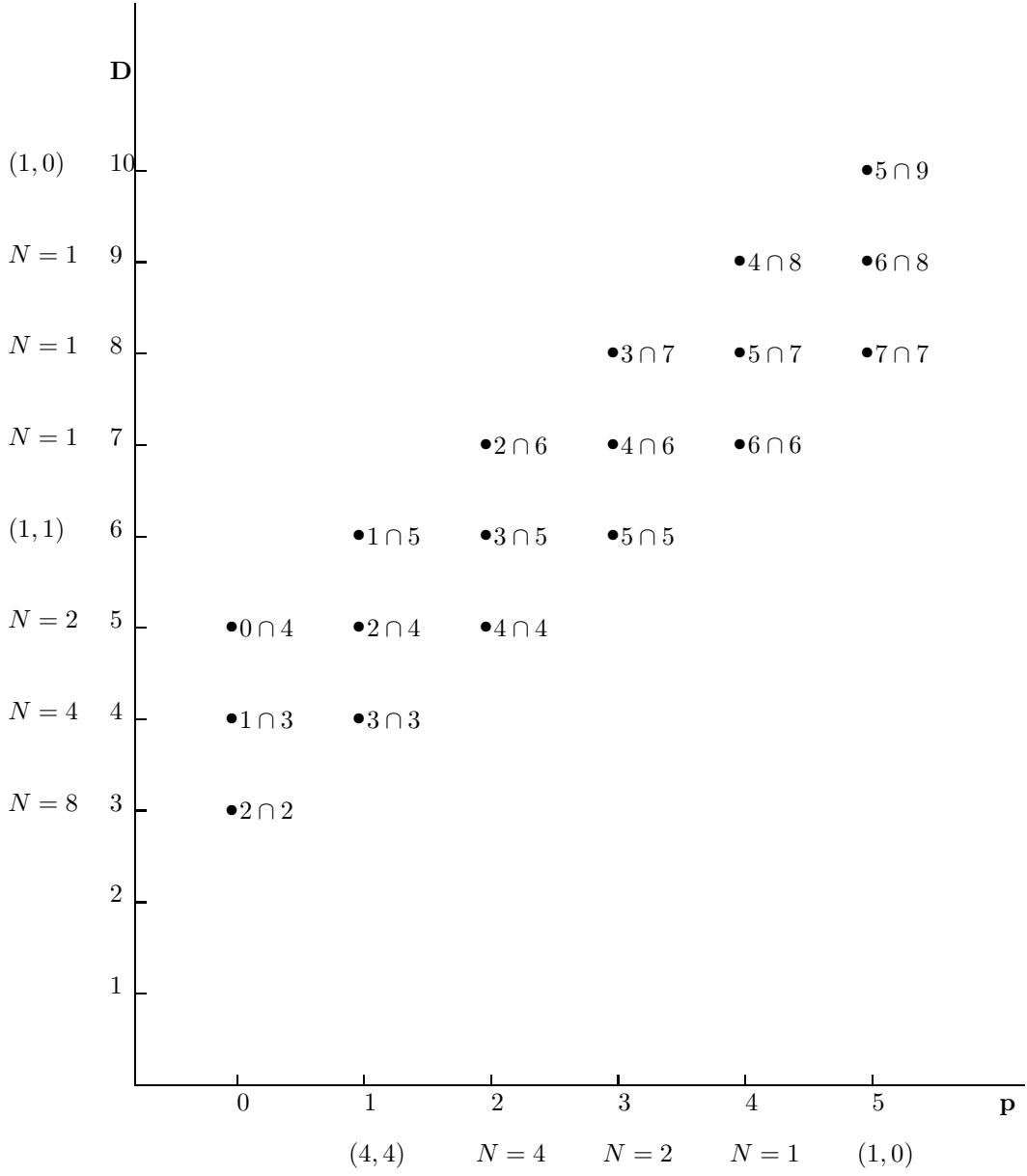
\begin{figure}[!ht]
\unitlength=.7mm
\leftline{
\begin{picture}(90,180)(-20,-20)
\put(-20,0){\line(1,0){175}}
\put(-20,0){\line(0,1){213}}
\put(-25,198){\makebox{${\bf D}$}}
\put(-25,180){\makebox{$10$}}
\put(-25,162){\makebox{$9$}}
\put(-25,144){\makebox{$8$}}
\put(-25,126){\makebox{$7$}}
\put(-25,108){\makebox{$6$}}
\put(-25,90){\makebox{$5$}}
\put(-25,72){\makebox{$4$}}
\put(-25,54){\makebox{$3$}}
\put(-25,36){\makebox{$2$}}
\put(-25,18){\makebox{$1$}}
\put(-45,180){\makebox{$(1,0)$}}
\put(-45,162){\makebox{$N=1$}}
\put(-45,144){\makebox{$N=1$}}
\put(-45,126){\makebox{$N=1$}}
\put(-45,108){\makebox{$(1,1)$}}
\put(-45,90){\makebox{$N=2$}}
\put(-45,72){\makebox{$N=4$}}
\put(-45,54){\makebox{$N=8$}}
\put(150,-5){\makebox{${\bf p}$}}
\put(125,-5){\makebox{$5$}}
\put(100,-5){\makebox{$4$}}
\put(75,-5){\makebox{$3$}}
\put(50,-5){\makebox{$2$}}
\put(25,-5){\makebox{$1$}}
\put(0,-5){\makebox{$0$}}
\put(122,-15){\makebox{$(1,0)$}}
\put(97,-15){\makebox{$N=1$}}
\put(72,-15){\makebox{$N=2$}}
\put(47,-15){\makebox{$N=4$}}
\put(22,-15){\makebox{$(4,4)$}}
\bull{125}{180}{5\cap 9}
\bull{100}{162}{4\cap 8}
\bull{75}{144}{3\cap 7}
\bull{50}{126}{2\cap 6}
\bull{25}{108}{1\cap 5}
\bull{0}{90}{0\cap 4}
\bull{125}{162}{6\cap 8}
\bull{100}{144}{5\cap 7}
\bull{75}{126}{4\cap 6}
\bull{50}{108}{3\cap 5}
\bull{25}{90}{2\cap 4}
\bull{0}{72}{1\cap 3}
\bull{125}{144}{7\cap 7}
\bull{100}{126}{6\cap 6}
\bull{75}{108}{5\cap 5}
\bull{50}{90}{4\cap 4}
\bull{25}{72}{3\cap 3}
\bull{0}{54}{2\cap 2}
\vgrid{0}\vgrid{25}\vgrid{50}\vgrid{75}\vgrid{100}\vgrid{125}
\hgrid{18}\hgrid{36}\hgrid{54}\hgrid{72}\hgrid{90}
\hgrid{108}\hgrid{126}\hgrid{144}\hgrid{162}\hgrid{180}
\end{picture}}\phantom{xxx}
\caption{Intersecting D-branes.
A D$q_1$ brane intersecting a D$q_2$ brane over
a $p$--brane is denoted by $q_1 \cap q_2=p$. The cases $q_1-q_2=4,2,0$
correspond to $q_1$ brane within $q_2$ brane, ending on $q_2$ brane or
intersecting with $q_2$ brane, respectively. In all these cases, $p$--brane is
viewed as moving in $D=q_2+2$ dimensional target space. Worldvolume and
target supersymmetries for these $p$-branes are shown along the $p$- and
$D$-axis, respectively. All the points shown in this Figure are related to each
other by $T$--duality transformations described briefly in the text.
\hfill}
\end{figure}}

\def\fd{
\begin{figure}[!ht]
\centerline{\psfig{figure=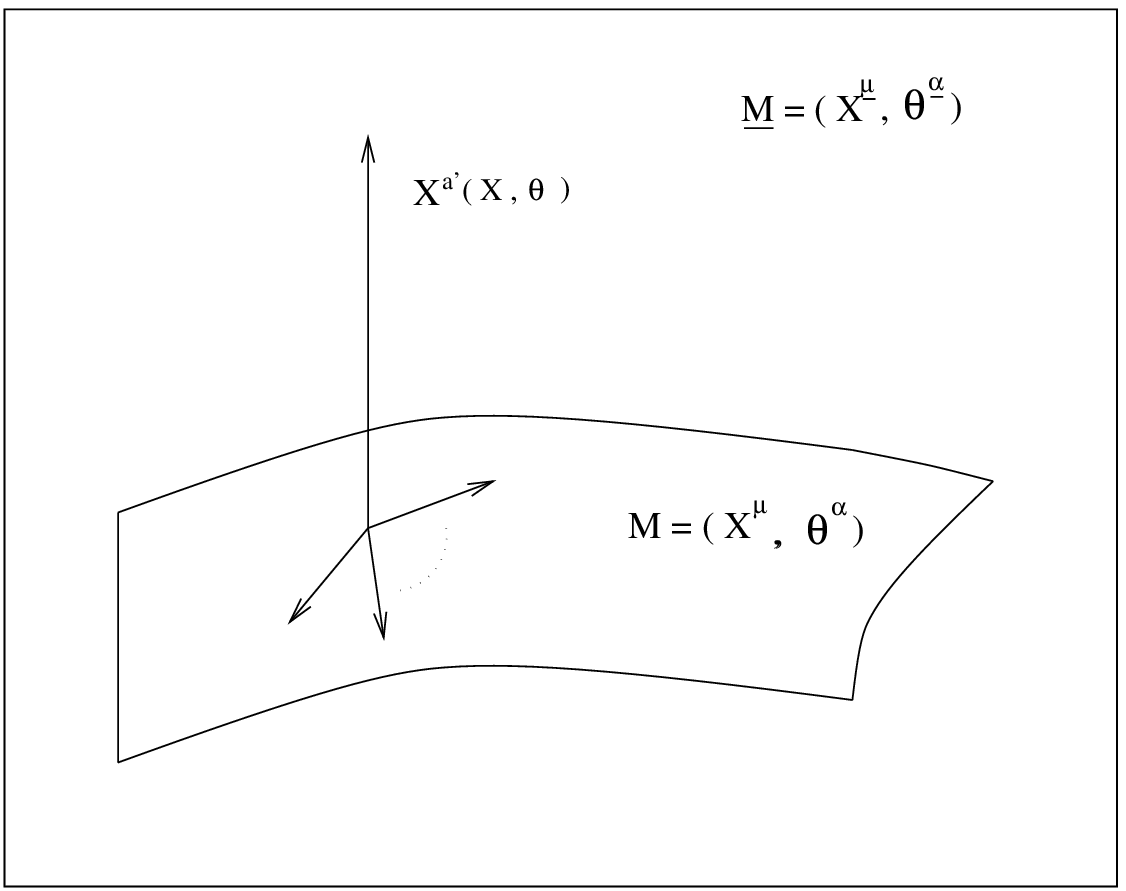}}
\caption{Superembedding of a world supersurface $M$ in a target superspace
${\unM}$. $X^{a'}$ indicates the transverse coordinates which are the
Goldstone superfields associated with the breaking of translations in ${\unM}$
to translations in $M$.}
\end{figure}}

\def\tp#1#2#3#4{$#1\cap#2\cap#3=#4$}
\def\ra{& $\longrightarrow$ &}
\def\da{$\downarrow$ && $\downarrow$ &&
$\downarrow$ \\}

\def\fe{
\begin{figure}[!ht]
\begin{center}
\begin{tabular}{ccccc}
\tp6220 \ra \tp7331 \ra \tp8442\\
\da
\tp5330 \ra \tp6441 \ra \tp7552\\
\da
\tp4420 \ra \tp5531 \ra \tp6642\\
\da
\tp3330 \ra \tp4441 \ra \tp5552 \\
\da
\tp4220 \ra \tp5331 \ra \tp6442 \\
\da
\tp5310 \ra  \tp6421 \ra \tp7532 \\
\da
\tp4400 \ra \tp5511 \ra \tp6622 \\
\da
\tp3310 \ra \tp4421 \ra \tp5532 \\
\da
\tp2220 \ra \tp3331 \ra \tp4442 \\
\end{tabular}
\end{center}
\caption{ The triple intersections of D-branes. The arrows
indicate various T-duality transformations.}
\end{figure}}

\def\ta{
\begin{table}[h]
\begin{center}
\begin{tabular}{|l|l|}
\hline
Branes  & Worldvolume Fields \\
\hline\hline
\multicolumn{2}{|c|}{D=11 M-Branes}\\
\hline\hline
pp-wave & \ \ \ $A_1,\, 8\times \phi$\\
\hline
M2-brane & \ \ \ $8\times \phi $\\
\hline
M5-brane & \ \ \ $A_2^+,\, 5\times\phi$ \\
\hline
KK monopole & \ \ \ $A_1,\, 3\times\phi$\\
\hline\hline
\multicolumn{2}{|c|}{D=10 Type IIA Branes}\\
\hline\hline
$Dp$-branes & \ \ \ $A_1,\, (9-p) \times \phi\ (\,p=0,2,4,6,8\,) $\\
\hline
Fundamental string &\ \ \ $8\times \phi$ \\
\hline
pp-wave & \ \ \ $A_1,\, 8\times \phi$\\
\hline
KK monopole & \ \ \ $A_1,\, 3\times \phi,\, S\sim A_4$\\
\hline
NS 5-brane & \ \ \ $A_2^+,\, 4\times \phi,\, S \sim A_4$\\
\hline\hline
\multicolumn{2}{|c|}{D=10 Type IIB Branes}\\
\hline\hline
$Dp$-branes & \ \ \ $A_1,\, (9-p) \times \phi\ (\,p=1,3,5,7,9\,) $\\
\hline
Fundamental string &\ \ \ $8\times \phi$ \\
\hline
pp-wave & \ \ \ $8\times \phi$ \\
\hline
NS 5-brane & \ \ \ $A_3,\, 4\times \phi$\\
\hline
KK monopole & \ \ \ $A_2^+,\, 3\times \phi,\, S\sim A_4,\, S'\sim A'_4$\\
\hline
\end{tabular}
\end{center}
\caption{Worldvolume bosonic fields. $A_q$ denotes a $q$-form potential,
$A_2^+$ is 2-form potential with self-dual field strength,
$S,S'$ are scalars which can be dualized to appropriate $q$ forms.
}
\label{ta}
\end{table}
}

\def\tb{
\begin{table}
{\footnotesize
\tabcolsep=1mm
\centerline{
\begin{tabular}{|c|cc|cccccccccccccc|}\hline
& & & & & & & & & & & & & & & &\\
{\large${}_{k}\backslash s$} & $0$ & \ns{$\ft12$} & $1$
& \ns{$\ft32$} & $2$ &
\ns{$\ft52$} &
$3$ & \ns{$\ft72$} & $4$ & $\cdots$ & $2s$ & $2s+$\ns{$\ft12$} & $2s+1$ &
$2s+$\ns{$\ft32$}
& $2s+2$ & $\cdots$ \\
& & & & & & & & & & & & & & & &\\ \hline
& & & & & & & & & & & & & & & &\\
$0$ & $35_{\smpl}\!+\!35_{\smm}$ & $56$ & $28$ & $8$ & $1$ &
& & & & & & & & & &\\
$1$ & $1\!+\!1$ & $8$ & $28$ & $56$ & $35_{\smpl}\!+\!35_{\smm}$
& $56$ & $28$ & $8$ & $1$ &
& & & & & & \\
$2$ & & & & & $1$ & $8$ & $28$ & $56$ & $35_{\smpl}\!+\!35_{\smm}$
& $\cdots$
& & & & & & \\
$\vdots$ & & & & & & & & & & & & & & & & \\
$s-1$ & & & & & & & & & & $\cdots$ & $1$ & & & & & \\
$s$ & & & & & & & & & & $\cdots$ & $35_{\smpl}\!+\!35_{\smm}$
& $56$ & $28$ & $8$ & $1$
& $\cdots$\\
$s+1$ & & & & & & & & & & & $1$ & $8$ & $28$ & $56$
& $35_{\smpl}\!+\!35_{\smm}$ & $\cdots$\\
$\vdots$ & & & & & & & & & & & & & & & & \\ \hline
\end{tabular}}}
\caption{{\small The $SO(3,2)\times SO(8)$ content of the symmetric
tensor product of two $N=8$ singletons. This product generates the
infinite spectrum of an $shs^E(8|4)$\ gauge field ($s\geq 1$) and the
spectrum of a finite number of additional matter fields ($s<1$).}}
\end{table}
}

\begin{document}

\title{ TOPICS IN M-THEORY
\footnote{Contribution to the Abdus Salam Memorial Meeting, 19-22 Nov 1997,
Trieste, Italy.} }

\author{ E. SEZGIN}

\address{Center for Theoretical Physics, Texas A\&M University,
College  Station, \\ TX 77843, USA}

\maketitle \abstracts {We give a brief history of the passage from
strings to branes and we review some aspects of the following topics in
$M$-theory: (a) an extended brane scan, (b) superembedding approach to
the dynamics of superbranes and (c) supermembranes in anti de Sitter
space, singletons and massless higher spin field theories.}


\section{Tribute to Abdus Salam}


Abdus Salam was a truly unique man with great achievements not only in
phycisc but also in promoting science in developing countries. His place
in the annals of science as one of the finest physicists in this century
is assured. He is certainly immortalized with his work on the
unification of electromagnetic and weak interactions. His achievements
in physics extend beyond this monumental work. He contributed many
important ideas in particle physics, covering important aspects of
renormalization theory, spontaneous symmetry breaking, grand unified
theories, superspace, string theory and supermembrane theory.

Abdus Salam's odyssey in physics began in earnest in 1950 when, after
having realized that he ``saidly lacked the sublime quality of
patience'' needed for conducting research in experimental particle
physics, he started to work under the guidance of Nicholas Kemmer, who
advised him to collaborate with Paul Matthews (who was completing his
PhD work at Cambridge University at the time) on renormalization of
meson theories. This marked the beginning of an amazing journey from the
pion-nucleon theory in 1950 to the marvelous discovery of the Standard
Model 17 years later. Abdus Salam has given a wonderful account of ``the
story of the short-lived rise of the pion-nucleon theory as the standard
model of 1950-51'' \cite{ir1}, and the ``story of the rise of chiral
symmetry, of spontenaous symmetry breaking and of electroweak
unification'', including the story of his interactions with Pauli,
Peierls, Ward, Weinberg, Glashow and others, in his Nobel Lecture of
1979 \cite{ir2}. Reading the account of the twists and turns encountered
in the remarkable odyssey which lead to the unification of electroweak
interactions, one feels the excitement of it and appreciates all the
more what the research in our discipline is really all about.

I first met Abdus Salam in 1981 in Trieste, when I joined the Abdus
Salam International Center for Theoretical Physics as a postdoctoral
fellow. This marked the beginning of a very enjoyable and fruitful
collaboration. I was privileged to have interacted with Abdus Salam for
more than a decade. I will always cherish this experience. It was
amazing how Abdus Salam treated a young post-doc that I was with so much
humility. When we completed our first paper \cite{ss1} he insisted that
I would put my name first. I had to argue vigorously to convince him to
put our names in alphabetical order.

During the 80's, we wrote a series of papers [6-19] and we edited a
reprint collection with commentaries on supergravities in diverse
dimensions \cite{dd}. Our work span topics in higher dimensional
Poincar\'e, anti de Sitter and conformal supergravities, their anomalies
and compactifications, string theory and supermembrane theory. Abdus
Salam was legendary in being open minded to new ideas. He embraced the
developments in the subject of supermembranes when not many others did.
He gave his full support for the supermembrane conference which was held
in Trieste in 1989. As far as I know, this was the first conference ever
to be held on membranes. The last papers we collaborated on
\cite{ss12,ss13} dealt with connections between membranes, singletons
and massless higher spin fields, which are among my favorites. I believe
that the full significance of the ideas put forward in those papers will
someday be better appreciated, in the process of discovering what
M-theory is. It was a joy to speculate about the tantalizing
brane-singleton-higher spin gauge theory connections in collaboration
with him.

Abdus Salam expressed his motivation in his research very humbly when he
said: ``I have spent my life working on two problems: first, to discover
the basic building blocks of matter; and secondly to discover the basic
forces among them'' \cite{ir3}. He was deeply religous man who realized
the limitations of science. He wrote: ``my own faith was predicated by
the timeless spiritual message of Islam, on matters on which physics is
silent, and will remain so'' \cite{ir3}. In the same article, he wrote:
`` the scientist of today knows when and where he is speculating; he
would claim no finality for the associated modes of thought". Such was
the humility and wisdom of the man.

Abdus Salam was a man with boundless energy and many creative ideas, not
only in physics but also in the process of promoting science in the
political domain. He travelled frequently all over the world and in
addition to his extremely productive research activities that resulted
with over 250 publications, he gave many speeches and he wrote several
articles on subjects other than physics. He had an amazing ability to
focus on the heart of matters at hand. He was always able to bear in
mind the big picture. He was very eloquent in his speeches and his
writing. Among many topics, ranging from the importance of transfering
science to developing nations to the interaction of science with
religion and society, he wrote with passion about the glorious era which
some of todays developing countries once had in science. He lamented the
decline of science in those countries and he was deeply disappointed
with the existing and ever widening gap between the developed and
developing countries. He wrote passionately about why most of the
developing countries need help in building up scienticic infrastructure
at all levels and why science transfer must accompany technology
transfer if the latter is to take root in those countries. He wrote:
``of the two passions of my life, the second has been to stress the
importance of ``science transfer'' for developing countries. After
building up the Theoretical Physics Department at Imperial College,
London, I have spent 20 years fighting the battle of stressing the
necessity of science transfer for developing countries'' \cite{ir4}. He
highlighted many important aspects of this and many other related
problems in brilliant speeches and essays which have been collected in a
book entitled ``Ideals and Realities'' \cite{ir}.

Abdus Salam made it a mission to himself to work towards the advancement
of science in developing countries. The immensely successful Abdus Salam
International Center for Theoretical Physics in Trieste (AS-ICTP) which
he founded in 1964 is a monument to his extraordinary achievement in
this sphere. The story of how AS-ICTP came to existence and how it
transformed to what it is today is an amazing one which can be
glimpsed from various essays that appeared in \citelow{ir}. Abdus Salam
was the primary driving force in this process from the very
begining. He always had brilliant ideas for how to expand the functions
of the Center and he saw to it that those ideas were actually put into
action \cite{ir}. Among many functions of the Center aimed at promoting
scientific activities in the developing countries perhaps the most
important one was to make it possible for the physicists from those
countries to visit the Center, and as he put it, to ``recharge their
intellectual batteries, work on research problems while at the Center,
and then return to their countries carrying a new line of work,
refreshed with new ideas and new interactions ''.

It should be pointed out that this success did not come easy. Abdus
Salam had to fight critical battles at times to ensure the continuation
and expansion of the Center. He worked incredibly hard to this end. He
had to make sacrifices, among which was the amount of time he could
devote to his belowed research activities. I remember once hopping into
a car which was taking him to the Venice airport for one of his frequent
trips, so that we could continue our physics discussion en route to the
airport. He was constantly at work in trying to ensure the success of
the Center and at the same time trying to carry out his research in
physics which he loved so much. With his relentless efforts, he
contributed to the advancement of science in the Third World in many
ways. Indeed, the legacy of Abdus Salam is not only his epoch-making
contribution to physics but also his brilliant contributions in building
up the scientific manpower in the developing nations.

Abdus Salam, a great man, brilliant, humanitarian, idealist, visionary,
articulate, eloquent and passionate, the man never at rest, is no longer
with us but he will always be remembered.


\section{ From Strings to Branes}


\subsection*{ A Brief History}


The notion of extended objects in the context of elementary particle
physics arose several years before the discovery of supersymmetry. The
most notable introduction of the idea is due to Dirac \cite{diracm}
who, in 1962, envisaged the possibility of the muon being an excited
state of a membrane in four dimension whose ground state corresponds to
an electron. With the discovery of supersymmetry in early 70's, the
physics of the extended objects took a remarkable turn, though for
nearly twenty years the focus was on the simplest extended object,
namely the string \cite{js1}.


It should be pointed out, however, that prior to the proposal of
Yoneya, Scherk and Schwarz in 1974 to interpret the dual models as
theories of elementary particles rather than hadrons \cite{js1}, there
were attempts \cite{n1,n2,n3,n4,n5} to generalize the dual models to
exhibit the four dimensional conformal $SO(4,2)$ symmetry. Inspired by
the connection between the dual resonance models and strings, a further
step was taken in \citelow{n4}(see also \citelow{n5}) to associate the
dual models possessing extended conformal symmetry with extended
objects (see \citelow{n6} for a review). In \citelow{n4}, the $(N-1)$
extra spatial dimensions were associated with a globally $SO(N,2)$
invariant theory, and these dimensions were interpreted as the orbital
degrees of freedom of $(N-1)$-dimensionally extended objects, related
to the ``dimension'' of the hadronic matter. In \citelow{n5}, the study
of the asymptotic behaviour of generalized dual model amplitudes led to
the consideration of $2k$ dimensional extended objects. In
\citelow{n6}, the intrinsic nonlinearity of Nambu-Goto type actions for
3-dimensional extended objects was recognized and the compactification
of a four dimensional worldvolume to a two dimensional worldsheet with
continuous internal symmetry was considered.


Turning to the story of super extended objects, to begin with, a
manifestly worldsheet supersymmetric Neveu-Schwarz-Ramond formulation
of string theory was discovered in 1971 (see \citelow{js1} for a
review). The considerations of anomalies led to the critical target
space dimension of $D=10$. This result, which can be derived in many
different ways, is one of the most amazing discoveries in physics. The
target space supersymmetry was not manifest in this formulation, but
this was remedied by Green and Schwarz \cite{gs1} who discovered just
such a formulation, though they had to sacrifice manifest worldsheet
supersymmetry. The Green-Schwarz superstring has a local fermionic
symmetry on the worldsheet known as $\kappa$-symmetry, which is
necessary for theory to make sense for several reasons. The Lorentz
covariant quantization of the Green-Schwarz string proved to be a
difficult problem, however, and consequently most of the work done in
string theory has been based on the Neveu-Schwarz-Ramond formulation.

It turned out that the extension of the superstring construction to higher
extended objects heavily favors the Green-Schwarz formalism, where one
constructs an action for the map from a bosonic $(p+1)$ dimensional worldvolume
to a target superspace. Significant progress towards the generalization of the
Green-Schwarz action to higher branes came after a better understanding of its
interpretation. A particularly useful such understanding was achieved in a
paper by Hughes and Polchinski \cite{hp} where the classical Green-Schwarz
superstring action in $D=4$ was understood as the effective low energy action
for a Nielsen-Olesen vortex solution of an $N=1, D=4$ supersymmetric Abelian
Higgs model such that the $N=1$ supersymmetry is broken down to $(2,0)$
supersymmetry in the $(1+1)$ dimensional worldsheet.  Soon after \cite{hlp},
the analog of this phenomenon was shown to arise in the context of a
three-brane solution of $(1,0)$ supersymmetric Yang-Mill theory in $D=6$, such
that, this time the $(1,0)$ supersymmetry is broken down to $N=1$ supersymmetry
on the worldvolume of the three-brane. The action was constructed for the
collective coordinates which form an $N=1$ scalar supermultiplet on the
three-brane worldvolume.

In 1987, inspired by these results, Bergshoeff, Townsend and the author
\cite{bst1} constructed an eleven dimensional supermembrane action. The target
space was taken to be a curved superspace, and the requirement of
$\k$--symmetry was shown to require the equations of motion of the eleven
dimensional supergravity!  Thus, connection was made between the eleven
dimensional supergravity which was invented  in its own right nearly a decade
before \cite{cjs} and supersymmetric extended objects. The action was
constructed directly without the knowledge of any membrane solution, which was
to be found years later \cite{ds}. It seemed to be a very natural extension of
the Green-Schwarz superstring action in $D=10$ to a supermembrane action in
$D=11$. Thus, it was very tempting to consider it as a candidate for the
description of a fundamental supermembrane theory that went beyond string
theory in a natural way.  While it was hoped that this passage from string to
membrane theory might have welcome consequences in solving some of the
outstanding unsolved problems of string theory, the theory was put forward
essentially because ``it was there''.  In other words, it was considered as a
logical possibility, primarily on the basis of symmetry considerations. It was
not invented out of pressing needs in physics based on paradoxes or anomalies,
with the possible exception of the desire to ``explain'' the existence of
$D=11$ supergravity.  In fact, once one comes to term with the basic idea of
transition from elementary particles to strings, then the passage from strings
to membranes is very natural.  Indeed, the higher than two dimensional extended
objects were also considered and their actions were constructed in
\citelow{bst1}.  A proper classification (with certain assumptions) was made
soon afterwards \cite{at} and it was found that the maximum target space
dimension allowed was $D=11$ and the maximum possible extension, $p$, of the
object was $p=5$. One of the important assumptions made was that the
worldvolume fields  always form scalar supermultiplets. It was a number of
years later that  branes which support other supermultiplets, most notably the
Maxwell supermultiplets in $(p+1)$ dimensions with $p=0,1,..,9$  and the 2-form
supermultiplets in $(5+1)$ dimensions were discovered.

The idea of a fundamental supermembrane in $D=11$ was pursued for couple
years after 1987 intensely by a number of authors. Primarily the following
issued were addressed: (a) spectrum and stability, (b) anomalies,  (c)
perturbation theory, (d) covariant quantization, (e) chirality/non-Abelian
internal symmetries (f) renormalizability (g) supermembrane in anti de
Sitter space and its relation to singleton field theory on the boundary of
AdS. We shall come back to these points later.

Considerably intense activities on supermembrane culminated in a Trieste
conference in 1989, devoted to the subject \cite{conf}. I believe this was the
first conference ever on supermembranes. The spectrum problem was emphasized
considerably, though other aspects of the supermembranes were also covered. The
spectrum issue appeared to be problematic due to the indications that ``the
supermembrane can grow hair'' without cost in energy, which seemed to imply a
continuum in the spectrum. No dent could be made in the quantization problem.
Despite the lack of covariant quantization scheme, and the lack of any
information about the consequences of the $\k$-symmetry at the quantum level,
the theory was widely considered to be nonrenormalizable. Moreover, it appeared
to be hopeless that the theory could ever produce a chiral spectrum by any
compactification scheme, for it seemed to be intrinsically nonchiral. It also
appeared that a realistic internal symmetry gauge groups could not be obtained.

There were, however, some results of promising nature by 1989 (see
\citelow{caracas} for an extensive list of references on (super)membranes
covering the period by mid 1990). To begin with, the particle \cite{bps87} and
string limits \cite{duff87} of the supermembrane were obtained. These limits
were sensible and they were suggestive of an important role for the
supermembrane to play in the larger scheme of things. The study the zero modes
of the supermembrane in the superparticle limit showed that the zero-mode
oscillators gave exactly the spectrum of massless states in $D=11$ supergravity
\cite{bps87}. Even more interesting was the result that a double dimensional
reduction of the $D=11$ supermembrane action yielded the type IIA superstring
action \cite{duff87}
\footnote{Interestingly, there exists a generalized Virasoro algebra for the
type IIA string which can be deduced from the algebra of worldvolume
diffeomorphisms and $\k$--symmetry of the $D=11$ supermembrane  \cite{ps}. }.
The significance of this result was not appreciated at the
time, partially due to the fact that the type II strings were  considered as
academic cases since they could not give rise to any promising physical picture
in $D=4$ by contrast with the heterotic string. It would have been nearly
impossible at the time to imagine that one day (in less than a decade, to be
more specific) the heterotic string would be related to eleven dimensional
supergravity and that all strings would be unified in an eleven dimensional
theory! See below.

In a related development, the $D=11$ supermembrane was quantized in the limit
in which the membrane was wrapped around a circle or torus \cite{duff88}.
This procedure brought into focus the study of the Kaluza-Klein modes of the
type IIA string as well. Again, the utility of the procedure of wrapping
membrane around compact spaces so as to examine the physics of the branes in
regions where they look like particles or strings, amenable to perturbation
theory was not fully appreciated until much later.

Another interesting development was the emergence of the area-preserving
diffeormorphisms, SDiff, of the supermembrane as a useful tool in the study
of the quantum theory \cite{hoppe} (for a generalization to volume preserving
diffeomorphisms of super $p$-branes, see \citelow{bst90}). The interesting
fact that the supermembrane Hamiltonian in a light-cone gauge turned into a
gauge Yang-Mills theory in $(0+1)$ dimensions (i.e. supersymmetric quantum
mechanics) with SDiff $\sim SU(\infty)$ as a gauge group was discovered. This
story too was to be appreciated later more fully, in the context of the
matrix model approach to M-theory \cite{mm}.

Last but not least, soon after the eleven dimensional supermembrane was
discovered, it was speculated \cite{duff1} that singletons, which are special
representations of the anti de Sitter group \cite{dirachs,ff3}, may play a role
in its description.  Soon after, it was conjectured \cite{bd1,nst} that a whole
class of AdS compactifications of supergravity theories may be closely related
to various singleton field theories. The singleton representations of the AdS
group are special in that they are ultra short and, strikingly, they admit no
Poincar\'e limit. Indeed, they can be realized in terms of fields that
propagate on the boundary of AdS \cite{ff3}.  The kinds of singleton field
theories studied back then were free field theories, and that raised the hope
that while supermembrane theory may seem to be nonrenormalizable in general, it
might miraculously be renormalizable in special backgrounds involving AdS
space, where it may be treated as a free singleton field theory. The recent
exciting developments in AdS/CFT correspondence \cite{malda,gkp,ew} is
reminiscent of these hopes, though the exact fashion in which this connection
has emerged certainly goes beyond what was known and imagined previously.  One
thing that was imagined before, and has not been materialized yet, is a
byproduct of the conjectured brane-singleton connection, namely the possible
occurrence of infinitely many massless higher spin fields in $AdS_4$ as two
supersingleton bound states in supermembrane theory!  This conjecture was put
forward in \citelow{ss12,ss13,bst00}. We will discuss this topic further in
section 3.

Despite these interesting developments, the quantization of the supermembrane
and higher superbranes remained to be an unsolved problem. As early as 1988,
however, aspects of branes as solitons or topological defects which break the
target space symmetries were revisited \cite{pktsol} and this proved to be a
very fruitful move. It was suggested \cite{pktsol} that all $p$--branes known
at the time should correspond to  solitonic solutions of certain
supersymmetric field theories (just as in the case of 3-brane of
\citelow{hlp}) or in the case of $10D$ strings and $11D$ supermembrane, they
should arise as solutions of appropriate {\it supergravities}. Few years
later, interesting and significant results started to appear in this
direction. In 1990, a string solution of $D=10$ supergravity was found
\cite{dab}. Soon after, a fivebrane solution of $D=10$ supergravity coupled
to Yang-Mills was discovered \cite{str1}. These results meant that string
theory contained solitonic objects in its spectrum which were nonperturbative
in nature. This implied a simple yet very important fact that supersymmetric
extended objects could not possibly be ignored any longer; they were simply
there and inevitable! The soliton fever carried over to eleven dimensions as
well. In 1991, the supermembrane soliton \cite{ds} and in 1992 the
superfivebrane soliton of $D=11$ supergravity were discovered. The  discovery
of the superfivebrane was somewhat surprising for it was not in the original
brane scan. Fascinating aspects of the superfivebranes have been discovered
since then and they, of course, occupy an important place in the big scheme
of things.

In the early 90's, a whole class of brane solutions in type II string theories
were also found. The study of type II branes eventually led to a remarkable
discovery by Polchinski in 1995 \cite{pol} that  the type II $p$-branes
carrying Ramond-Ramond charges can be understood as Dirichlet branes, or
D-branes for short. These are $p$ dimensional surfaces on which an open string
can end. Thus, it became possible to study the dynamics of at least special
kinds of $p$-branes (at weak coupling limit) by studying the (perturbative)
dynamics of  an open string! The ``D-brane technology'' developed rapidly (see
\citelow{pol2} for a review) and it provided an important tool for studying the
dynamics of intricate brane configurations, leading to discoveries of novel
physical phenomena and to breakthroughs in the study of long standing problems
in black hole physics.

Concomitant to these developments, and closely related to them, were the
important discoveries in the arena of duality symmetries of string theory.
Remarkable results on T-duality symmetries relating strings in backgrounds
involving a circle of radius $R$ and $1/R$ and $S$-duality transformations
interchanging strong and weak coupling limits started to accumulate with
increasing rate. This is a vast subject, even a brief review of which would
take us beyond the scope of this brief introduction. We refer the reader to
the excellent reviews of this subject existing in the literature; see for
example, [129-138].

Further studies in strong-weak coupling dualities led to major developments in
late 94 and early 95 which finally put the string-membrane connection on a firm
footing. Firstly, it was observed that the soliton spectrum of the
compactified $D=11$ supergravity  agreed with that of compactified type IIA
string by the inclusion of the wrapping modes of the supermembrane and
superfivebrane  and by taking into account the wrapping  modes of the type IIA
D-branes carrying Ramond-Ramond charges \cite{ht1}. Next, it was argued that
the $D0$--branes of type IIA string correspond to the Kaluza-Klein modes of
$D=11$ supergravity \cite{pkt1}. These were tantalizing new hints for a deep
connection between type IIA string and the eleven dimensional supermembrane
that went beyond the  connection based on double dimensional reduction, because
it was not necessary to consider only the zero modes. Finally, Witten in his
celebrated paper \cite{ew95} argued convincingly that the strong coupling limit
of type IIA string theory {\it is} the $D=11$ supergravity! Furthermore, the
ensuing developments showed that all string that all string theories in $D=10$
were unified via duality relations involving an eleven dimensional origin in
one way or another! The view started to develop that there exists an
intrinsically nonperturbative and  quantum consistent eleven dimensional theory
with the properties that (a) it can be approximated by $D=11$ supergravity at
low energies, (b) it contains the supermembrane and superfivebrane in its
spectrum and (c) when expanded perturbatively in different coupling constants, it
gives different perturbative theories, which can be superstrings or
superparticles. This mysterious  theory was named (upon a suggestion by Witten)
the M-theory \cite{jhs0}. Work on M-theory continues with rapid pace (see, for
example, \citelow{js2,pktr,sen} for technically detailed reviews) and striking
new results have emerged from the studies of M-theory in anti de Sitter
background \cite{malda,gkp,ew}.

In the light of these developments, it is interesting now to revisit the
questions that arose in the late eighties in the context of the eleven
dimensional supermembrane, which were mentioned above. The problem of
chirality and non-Abelian internal symmetries found a remarkable answer with
the discovery made by Horava and Witten \cite{hw} that M-theory compactified
on an interval leads to the $E_8\times E_8$ heterotic string! This phenomenon
provides a surprisingly simple and elegant answer to the question of how to
obtain a chiral theory starting from eleven dimensions. As for the problems
associated with the quantization of the supermembrane, there is no solution
in sight (not yet!)

\vspace{-0.25cm}

\subsection*{M Theory-Supermembrane Connection}

Notwithstanding the presently unsolved problem of how to quantize the
supermembrane (perturbatively or nonperturbatively) in eleven dimensions, it is
tempting to pose the following question: Is it conceivable that M-theory is
nothing but the eleven dimensional supermembrane theory? According to
\citelow{jhs3}, ``most experts now believe that M-theory cannot be defined as a
supermembrane theory''. While we are not aware of all the arguments leading to
this assertion, some of them go as follows: (a) The supermembrane action is
only meant to describe a macroscopic object in M-theory and therefore one should
not even attempt to quantize the supermembrane. (b) The fundamental and
solitonic supermembranes should be identified \cite{ht1}. The latter has a
finite core due to its horizon \cite{dgt}. Since the known supermembrane
action does not take into account this classical structure of the membrane,  it
is not an appropriate starting point for quantization \cite{pkt1,pktdem}. (c)
As far as the perturbative formulation goes, there is no suitable perturbative
expansion parameter (assuming that the theory is not compactified) to order the
spacetime amplitudes and moreover the worldvolume perturbation theory is
non-renormalizable.

Despite all these considerations, the eleven dimensional supermembrane theory
seems to be the only theory that goes beyond $D=11$ supergravity and which does
incorporate supergravity equations of motion. Indeed, it does so already at the
classical level, thanks to the power of $\k$--symmetry. The theory goes beyond
$D=11$ supergravity because we know that the quantization of the supermembrane
collapsed to a point yields the $D=11$ supergravity spectrum, and that the
wrapping the supermembrane around a circle gives type IIA string theory, a
perturbative treatment of which yields an infinite set of higher derivative
corrections to type IIA supergravity. It is natural to expect a supermembrane
origin of these corrections.

In this context, let us recall that while the $\k$--symmetry of the
supermembrane uniquely leads to the $D=11$ supergravity equations of motion
\cite{bst1}, there is one correction to these equations \cite{dlm} that has
been motivated by the considerations of sigma model anomalies in $M5$--brane
and one-loop effects in type IIA string \cite{ewv}, and takes the form
$C_3\wedge X_8$,  where $C_3$ is the 3-from potential in $D=11$ and $X_8$ is an
8-form made out of the Riemann curvature.  Supersymmetrization of this term
implies an infinite number of terms in the action, just as in the type  IIA
theory in $D=10$. Some aspects of these terms have been deduced from one-loop
effects in $D=11$ supergravity, but this cannot be the full story
\footnote{
It is nonetheless an amazing  fact that these
one-loop effects are capable of producing nonperturbative effects in  type II
string theory \cite{mg}.
}.

What then is the principle which governs the corrections to $D=11$
supergravity? Since we no longer have the benefit of worldsheet
superconformal invariance in curved background that helps in answering a
similar question in $D=10$, we have to find a new principle here. Local
supersymmetry is very helpful, but we need more than that, based on the
lessons learnt from string theory. We mentioned above that the $\k$--symmetry
of the supermembrane was very restrictive by giving the usual $D=11$
supergravity equations. Perhaps one should re-examine the issue of
$\k$-symmetry by allowing more general superspace than the standard $D=11$
Poincar\'e superspace.

To have a control over the higher derivative corrections in $D=11$, it is very
useful to work in superspace. One possible approach is to modify the
superspace Bianchi identity as follows

\be
dH_7= G_4 \wedge G_4 +X_8\ , \la{dh}
\ee

where the $G_4=dC_3$ and $H_7$ is a 7-form whose purely bosonic components are
Hodge dual to those of  $G_4$. The $\th=0$ component of this equation has been
extensively discussed in \citelow{dlm}, but not much is  known about the
superspace consequences of the full equation, essentially due to its
complexity. It is possible that the solution requires the modification of the
standard  supergeometry by introducing the 2nd and/or 5th rank $\C$--matrix
terms in the constraint for the dimension zero supertorsion components
$T^a_{\a\b}$ \cite{psh,bn}. It would be very interesting  to derive the
corrections  to the minimal superspace geometry from the $\kappa$--symmetry
considerations, or better yet, from the formalism of superembeddings, which
will be summarized in section 4. It is clear that any modification of the
standard $D=11$ supermembrane action, or equations of motion, would be very
interesting and it would  effect the discussion of just what is the role of the
supermembrane in M-theory.

In the above discussion we omitted the superfivebrane. It is natural to expect
a sort of duality relationship between supermembranes and superfivebranes (see
\citelow{duff96} for a discussion of the membrane/fivebrane duality in $D=11$).
Moreover, it is well known that an open supermembrane can end on a
superfivebrane in $D=11$. Thus, the connection between $M2$--branes and
$M5$--branes in $D=11$ is similar (in some respects) to the connection between
open strings and $Dp$--branes in $D=10$. In fact, it has been shown \cite{cs}
that the superfivebrane equations of motion follow from the $\k$--symmetry of
an open supermembrane ending on a superfivebrane!

We conclude this introduction by emphasizing the importance of exploring the
consequences of the M-theory unification at the level of interactions. Much of
the work done so far understandably has dealt with spectral issues and this has
been very beneficial. However, at some point several problems   about
interactions need to be probed more deeply. Some encouraging  results have
already emerged. Interesting connections between  the string interactions in
$D=10$ and membrane interactions in $D=11$ have been noted \cite{ah1}. We
already mentioned the one-loop effects in $D=11$ supergravity giving rise to
nonperturbative information about type IIA string amplitudes \cite{mg}. Another
example is the deduction of the quantum Seiberg-Witten effective action for
$N=2$ supersymmetric Yang-Mills theory \cite{sw1} from the classical
M-fivebrane equations  of motion with $N$ three branes moving in its
worldvolume \cite{hlw}
\footnote{
It is rather amusing to see that the Planck constant emerges as a combination
of certain integration constants arising in the course of solving the
three-brane equations \cite{hlw}.
}.

In the next section we will discuss an extended brane scan covering  various kinds of
branes that have emerged until now. The emphasis will be on the amount of
supersymmetry breaking by the branes. The purpose of this section is to give a feel
for the panorama of branes in M-theory and their properties. The superembedding
theory plays a significant role not only in their classifications but also in the
description of their dynamics. With this motivation in mind, section 4 is devoted to
a summary of the basic ideas behind the superembedding theory.  When the target space
is taken to be a supercoset involving anti de Sitter space, remarkable things happen,
as it has become abundantly clear with recent exciting developments. In section 5, we
summarize some aspects of the connections between  supermembranes, singletons and
higher spin gauge theory.

\vspace{-0.25cm}


\section{Types of Branes}


\subsection*{ A Minimal Brane Scan (The Scalar Branes)}


Until the discovery of the superfivebrane solution of the $D=11$
supergravity in 1992 \cite{g}, the types of branes that were known were
rather limited in number. Some of them were already anticipated in
\citelow{bst1} and classified properly in \citelow{at}. The result
is reproduced in Table 1. For uniformity in the nature of the scan, we
have left out the type IIB, type I and heterotic strings. The main
characteristic of all the branes occurring in this scan is that they all
support a scalar multiplet with $1,2,4,8$ scalars and matching spin
$1/2$ fermionic degrees of freedom. The branes in this scan fall into
four series: The octonionic, quaternionic, complex and real series with
co-dimensions 8,4,2,1 embeddings, respectively. All the members of a
given series can be obtained form the one that occupies the highest
dimension by the process of double dimensional reduction \cite{duff87}. All
branes in this scan for $p >1$ have minimal possible target space
supersymmetry.

The brane scan shown in Figure 1 was originally derived from the requirement of
$\k$-symmetry of their actions \cite{bst1}. This requirement amounted to the
existence of suitable Wess-Zumino terms which was possible whenever a closed super
$(p+2)$ in target superspace existed. This in turn meant finding the values of the
pairs $(p,D)$ for which the following $\C$-matrix identity is satisfied

\be
\C_{\m (\a\b}\,
\C^{\m\n_1\cdots \n_{p-1}}_{\c\d)}=0\ ,
\ee

where $\m=0,1,...,D-1$ and $\a$ labels a minimal dimensional spinor in
$D$-dimensions. There is a simple alternative way to deduce the same brane
scan. Indeed, using the rule

\be
D=(p+1)+n_S\ , \la{scan}
\ee

where $D$ is the bosonic dimension of the target space, $p$ is the spatial
brane dimension and $n_S$ is the number of real scalars in the scalar
multiplet, together with the knowledge of which supermultiplets exist in
diverse dimensions, the brane scan shown in Table 1 can easily be reproduced.


Of the branes for $p>1$, only the supermembrane in $D=11$ attracted the most
interest for some time. There was a modest attempt to try to rule out the
``other branes'' on various grounds \cite{bps87,ms,bars,bp} though, primarily
on the basis Lorentz anomaly considerations. At the time, intersecting branes
\cite{pktint} were not known and all the scalar branes were considered in their
own right. The constraints implied by the $\kappa$--symmetry of their actions
have been determined long ago \cite{bst1}. The consequences of these
constraints have not been studied so far except for the cases of strings in
$D=3,4,6,10$ {\cite{bst346} and the supermembrane in $D=11$ \cite{bst1}. In the
case of fivebrane in $D=10$, one can check that the dual formulation of $(1,0)$
supergravity is consistent with the $\kappa$-symmetry constraints, though it is
considerably more difficult to show that it is uniquely implied by these
constraints. As for the solution of the $\kappa$--symmetry constraints for the
remaining branes that occur in the old brane scan (see Figure 1), we expect
that the dimensional reduction of the dual $(1,0)$ supergravity in $D=10$,
which contains a 6-form potential, followed by a truncation of the resulting
vector multiplets,  provides a solution.

While the old $p$--branes we have been discussing may emerge in the physics of
intersecting branes, it is still interesting to determine if they can describe
consistent brane theories formulated in subcritical $D<10$ dimensional
spacetimes.  If so, these branes might correspond to an interesting class of
M-theory limits in which the $(10-D)$ or $(11-D)$ extra dimensions decouple in
a rather drastic way.  In fact, this is somewhat reminiscent of the situation
arising in the so called ``little m'' theories \cite{mandm}.  However, in the
latter case the target space theory typically involves the Yang-Mills
supermultiplet but not the supergravity fields. For example, the little
$m$-theory in $D=7$ involves only the super Yang-Mills theory.  This is to be
contrasted with the supermembrane in $D=7$ which arises in the old brane scan
(see Figure 1), where the target space theory naturally involves the $N=1, D=7$
supergravity but not super Yang-Mills.  In fact, this raises the interesting
question of whether one can couple supergravity plus Yang-Mills system to the
supermembrane in $D=7$ that arises in the old brane scan.  If such a coupling
exists, it would be reasonable to expect a limit in which supergravity is
decoupled.

\fa

Another aspect of the branes listed in Figure 1 which might be worthwhile to
study is their anomalies; namely the gravitational anomalies in the target
field theories as well as the $\kappa$--symmetry and global anomalies on the
worldvolume. In doing so, one immediately rules out the fivebrane in $D=10$ on
the basis of its incurable gravitational anomaly. However, many of the lower
dimensional branes, in particular those for which the target space is
odd-dimensional such anomalies do not arise. For example, the supermembrane in
$D=7$ has odd dimensional target as well as odd dimensional worldvolume, and
consequently, the pertinent anomalies are the global ones. Indeed, such
anomalies have been studied by Witten \cite{ewflux} in the case of the $D=11$
supermembrane, and it was found that a mild restriction arises on the topology
of spacetime and the possible configuration which the membrane Kalb-Ramond
field may assume. With similar conditions satisfied, we expect that the
supermembrane in $D=7$ is anomaly free, though we do not know at present how
this brane might possibly arise in M theory.

\vspace{-0.25cm}


\subsection*{  M-Branes and D-Branes }


With the discovery of the superfivebrane, the novel situation in which
the worldvolume supported a multiplet other than scalar multiplet emerged.
Indeed, it was found that the $5+1$ dimensional worldvolume theory was that of
$(2,0)$ tensor multiplet, containing a two-form potential with self-dual field
strength, giving rise to $3$ on-shell degrees of freedom, five scalars and $8$
on-shell fermi degrees of freedom. The rule \eq{scan} still holds. It was
speculated later that there should be an analog of this brane with $(1,0)$
tensor multiplet that contains a single scalar, in addition to the a self-dual
tensor and $4$ on-shell fermions. The single scalar suggests a seven
dimensional target space. This model has been analyzed in considerable detail
in \citelow{b1}.

In 1985, the $D$-branes were discovered \cite{pol}. These are branes on which
open strings can end. The worldvolume dynamics of these branes is  described by
vector multiplets. Considering the maximally symmetric Maxwell multiplets in
various dimensions, i.e. those with $8$ bose and $8$ fermi on-shell degrees of
freedom and using the rule \eq{scan}, one finds they all imply $D=10$ target
space with type II supersymmetry. Allowing nonmaximal vector supermultiplets on
the worldvolume gives rise to a revised brane scan \cite{dl1} and  applying
\eq{scan} one finds $D=3,4,6$ dimensional target spaces.

At this point, it is tempting to contemplate a classification of all
possible branes by classifying all possible globally supersymmetric
multiplets in dimensions $p+1 \le 10$. There are some complicating
factors, however.

Firstly, there is the possibility of dualizing a member of the worldvolume
supermultiplet, say a $q$-form potential, to a $p-q-1$ form potential.  Indeed,
one may consider the dualization of a number of forms at the same time. While
this may be a simple matter at the free field theory level, it is considerably
more difficult for branes where the worldvolume multiplets are self-interacting
in a highly nonlinear fashion.  In fact, the dualization of forms on the
worldvolume is intimately connected with the $S, T$ or more general  duality
transformations.  Consequently, there is the additional question of which
branes should be considered as the fundamental ones from which all others can be
derived by one duality transformation or by  Kaluza-Klein type reductions of
the target space and/or the  worldvolume.

Secondly, it is possible that the description of the worldvolume theory
involves more than one supermultiplet. The simplest example of this is
the heterotic string where the worldsheet contains scalar multiplets in
the left-moving sector and heterotic fermions or bosons, which are
supersymmetry singlets, in the right-moving sector. The fact that
worldsheet supersymmetry allows supersymmetric singlets is special to
$1$ and $2$ dimensions, and it cannot occur in higher than two
dimensions. Nonetheless, focusing our attention on the fact that the
heterotic string is described by two distinct multiplets on the
worldsheet, we tentatively refer all branes which support more than one
irreducible supermultiplet as heterotic branes.

Given the complications just described, the classification of branes becomes a
rather nontrivial task. However, one may consider an alternative scheme in
which one focuses on the amount of supersymmetry breaking due to the embedding
of  superbranes in a given target superspace rather than emphasizing the
worldvolume supermultiplet. This approach especially becomes powerful
if one considers both the target space as well as the worldvolume to be
superspaces. Thus, one considers the embedding theory of supersurfaces, which
turns out to be a geometrical and natural framework, not only for classifying
the superbranes, but also for providing the manifestly worldvolume and target
space supersymmetric dynamical equations. Indeed, the problem of describing the
superfivebrane equations of motion was solved for the first time by using this
formalism \cite{hs1,hs2}. The main criteria in the theory  superembedding
theory from the physical point of view is that the basic equations which
describe the superembedding lead to sensible equations of motion for the
worldvolume fields. This can be typically checked in a reasonably
straightforward way at the linearized level, at least for a large class of
superbranes.

The classification of all possible branes is still a formidable task despite
the universal nature of the superembedding approach.  A further  complicating
factor is that the geometry and topology of the spaces involved can
mathematically become rather complicated.  For example, the full actions for
intersecting branes is yet to be constructed in any approach. Nonetheless the
existence of a large class of branes has been deduced from the study of brane
solutions to supergravity theories, or from the study of the possible brane
charges in supersymmetry algebras, or from the analysis of the linearized
embedding equations.  Putting together a variety of information available on
the nature of existing and predicted types of branes, they can be classified
according to the amount of supersymmetry they preserve.  For the purposes of
the proposed scan, we will assume that the maximum real dimension of
supersymmetry is $32$ and we shall furthermore consider flat target superspaces
with Lorentzian signature.

\vspace{-0.25cm}


\subsection*{ $32 \rightarrow 16$  Branes}


Assuming Lorentzian signature, the maximum dimension in
which real $32$ symmetries can occur is $D=11$. In $D=11$ the target
superspace has (even|odd) dimensions $(11|32)$. Embedding a $(3|16)$
dimensional super worldsurface gives the supermembrane, and embedding a
$(6|16,0)$ dimensional super worldsurface gives the superfivebrane. The
former is a co-dimension $8$ embedding, and the later is a co-dimension
$4$ embedding. In the latter case, the notation $(16,0)$ means that
there are $16$ real left-handed spinors and no right-handed spinors.
This means $(2,0)$ supersymmetry, since the minimum real dimension of a
spinor in $5+1$ dimensions is $8$. Such hybrid superspaces can occur in
$2\,mod\,4$ dimensions (assuming Lorentzian signature).

The supermembrane worldvolume multiplet is an $N=8$ scalar multiplet,
which has $8$ real scalars and the superfivebrane worldvolume multiplet
is the tensor multiplet of $(2,0)$ supersymmetry which has $5$ real
scalars and a 2-form potential with self-dual field strength. Thus, the
relation \eq{scan} holds in both cases.

In $D=11$ there are two other ``superbranes'' which preserve half
supersymmetry, but for which the relation \eq{scan} does not hold. These are
the pp-waves, which can be considered in some sense as $1$-branes and the
Kaluza-Klein monopole, which can be viewed in a certain sense as $6$-branes.
The existence of $9$-branes has also been speculated briefly in \citelow{hs1}
and  in some more detail in \citelow{hull}. Sometimes the boundaries of the
$D=11$ spacetime arising in the Horava-Witten configuration \cite{hw} is also
referred to as a $M9$-brane. See \citelow{ber3} for a recent discussion of
various  aspects of $M9$ branes.

In all these cases, the formula \eq{scan} breaks down, and one finds $8$ scalars
for the pp-wave, $3$ scalars for the KK monopole and no scalars for the $9$-brane,
as opposed to $9,4,1$ scalars, respectively. This is essentially due to the fact
that the transverse space lacks the isometries of $R^9, R^4, R^1$, respectively;
for example, in the case of KK monopole, the transverse space is a Taub-Nut space.
A detailed discussion of these branes can be found, for example, in \citelow{hull}
where they have been called the $G$-branes ($G$ standing for gravitational).

In classifying the superbranes in $D=10$, one should take into account the
ordinary (vertical) dimensional reduction, or (diagonal) double dimensional of
the branes in $D=11$. Moreover, one may take the eleventh dimension to be $S^1$
or $S^1/Z^2$. This would generate a set of branes known to exist in $D=10$ (See
Figure 2).

In $D=10$ we can embed the $(p+1|16)$ dimensional $Dp$-brane
worldsurfaces for $p=1,3,5,7,9$ in $(10|16,16)$ superspaces, where
$(16,16)$ denotes the $16$ left-handed and $16$ right handed
Majorana-Weyl spinor coordinates, associated with type IIA $(1,1)$
supersymmetry. Similarly, the $(p+1|16)$ dimensional $Dp$-brane
worldsurfaces for $p=0,2,4,6,8$ can be embedded in the $(10|32,0)$
superspace associated with type IIB $(1,0)$ supersymmetry. The
worldvolume supermultiplets for the $Dp$-branes are the maximally
supersymmetric vector multiplets in $p+1$ dimensions, which have a single
vector and $(9-p)$ scalars in the bosonic sector. Thus, the relation
\eq{scan} holds for all these branes.

\fb

In $D=10$, there also exist the fundamental strings, NS$5$-branes, pp-waves and
Kaluza-Klein monopoles (which may be viewed as $5$-branes), both in type IIA and
type IIB superspaces.  These are denoted by $1_F$, $5_S$, $1_W$ and $5_{KK}$,
respectively, in Figure 2. The $Dp$ branes for $p=-1,7,8,9$ are somewhat special
cases which are not shown in this Figure.  The type IIA/B branes are related to
each other by T- and S- dualities as shown in Figure 2.  The type IIA theory
compactified on a circle of radius $R$ is $T$-dual to the type IIB theory
compactified on a circle of radius $1/R$.  The $SL(2,Z)$ $S$-duality
transformation, on the other hand, is a strong-weak coupling symmetry operative
in type IIB theory:  the $(1_F, 1_D)$ and $(5_S,5_D)$ form a doublet and $3_D$ is
singlet under this symmetry.  The action of the $S$-duality on the type IIB
$D7$-- and $D9$--branes is more involved \citelow{ggp,hull,mo}. For a more
detailed version of Figure 2 which summarizes almost all the known dualities
between the type II branes, including the $D7$-- and $D9$--branes, see
\citelow{mo}. For the S-duality involving type IIB KK-monopoles, see
\citelow{eyras2}.

The worldvolume field content of all these branes have been obtained in
some cases speculated in a number of references. We refer to \citelow{hull}
where an extensive discussion and references to earlier work can
be found. For reader's convenience and following \citelow{hull} and
\citelow{ber2}, the $D=11$ and type II brane worldvolume contents are listed
in Table 1.\\

\ta

\subsection*{ $16 \rightarrow 8$ Branes/ Doubly Intersecting Branes }


The maximum dimensional target space admitting $16$ real component spinor is
$D=10$. There is only one case to consider here, which is the $(1,0)$
supersymmetric $(10|16,0)$ superspace (or any of its dimensional reductions).
We can embed a super $5$-brane with $(6|8,0)$ world supersurface. This gives a
hypermultiplet of $(1,0)$ supersymmetry on the worldvolume and this is the old
super $5$-brane which was considered long ago in its own right, prior to the
discovery of intersecting branes. The target supergravity theory is the dual
formulation of $(1,0)$ supergravity, which has fatal gravitational anomalies
(see \citelow{dd} for a review). There is an alternative way to interpret this
embedding, however. It can be viewed as a $D5$-brane within a $D9$-brane of
type IIB theory. This is shown as the point $5\cap 9=5$ in Figure 3. It is
important to note that both branes are embedded in type IIB superspace which
has $32$ real supersymmetry, though the $5$-brane residing inside the
$9$-brane possesses only $8$ real supersymmetries. This is known as $1/4$
supersymmetry preserving brane.

\fc

We also emphasize that the supergravity theory in the target of this
$5$-brane is induced supergravity where all the members of the
supergravity multiplet are composites of the type IIB supergravity
theory. Since type IIB theory is free from anomalies, the $D5$-brane
residing inside the $D9$-brane is presumably anomaly free as well.
Starting from the fivebrane inside the ninebrane, one can perform two kinds of
$T$-duality transformation which generates all the intersecting branes shown
in Fig. 3. Starting from an intersecting pair $p\cap q$, a vertical move
corresponds to a $T$ duality transformation over one of the transverse
directions of the $q$ brane. An oblique move corresponds to $T$-duality over
one of $q$-brane the worldvolume directions. The details of such
transformations to obtain one brane solution from the other is treated in a
number of papers; see for example, \citelow{berint,gaunt}.


Intersecting branes involving $NS$ branes can be obtained by utilizing combined
$S/T$-duality transformations. Not including the cases involving the pp-waves
and KK monopoles one thus finds

\bea
Dp \cap 1_F &=&  0\ ,\qquad \quad \quad  \ \ p=1,2,...,9\ ,
\nn\\
Dp \cap 5_S &=& (p-1)\ , \qquad p=1,2,...,6\ ,
\nn\\
1_F \cap 5_S &=& 1\ ,
\nn\\
5_S \cap 5_S &=& 3\ .  \la{op}
\eea

For completeness, we also list the by now well known double intersections of
M-branes:

\be
5\cap 5=3\ ,\quad\quad 2 \cap 5= 1\ , \quad\quad 2 \cap 2 =0\ .
\ee

The relative and overall transverse dimensions for these intersection are
$(4,3)$, $(5,4)$ and $(4,6)$, respectively.

\vspace{-0.25cm}


\subsection*{ $8 \rightarrow 4$ Branes/ Triply Intersecting Branes }


The maximum dimensional target space admitting $8$ real component spinor is the
$(1,0)$ superspace in $D=6$ (or any of its dimensional reductions). We can embed a
superstring with $(2|4,0)$ super worldsheet, or a super 3-brane with  $(4|4)$ world
supersurface in this target superspace. This gives the  $(4,0)$ supersymmetric
hypermultiplet on the string worldsheet and the $N=2$ scalar multiplet on the 3-brane
worldvolume. Considered as scalar branes, these are shown in Table 1. An alternative
way to interpret the above  embeddings is to view them as triple intersections of
suitable M- or D-branes. the multi-intersections of all known branes have been
extensively studied in a number of papers. The most extensive classification known
to us can be  found in \citelow{berint}. Here, we shall be content with the
reproduction of the list for triply intersecting M- and D-branes.

\fe

To begin with, the complete list \cite{ber2} of triple intersections of
D-branes is given in Figure 4. The arrows indicate various T-duality
transformations whose precise nature is spelled out in \citelow{ber2}.  The
common property of these intersections is that the relative transverse
dimension for any pair is always $4$. The overall transverse dimension is
$(3,2,1)$ for the triple intersections over $(0,1,2)$--branes, respectively.

Finally, the triple intersections of M-branes are given by

\bea
&& 5\cap 5\cap 5 =3\ , \quad\quad 5\cap 5\cap 5 = 2\ ,\quad\quad
5\cap 5\cap 5 = 0\ , \nn
\w2
&&5\cap 5\cap 5 = 1\ , \quad\quad 2\cap 5\cap 5 = 1\ ,\quad\quad
2\cap 2\cap 5 = 0\ , \nn
\w2
&&2\cap 2\cap 2 = 0\ .
\eea

The relative transverse dimension for any pair in all of these intersections is
always $4$ or $5$ (see, for example, \citelow{berint}).\\

\vspace{-0.25cm}

\section{Superbrane Dynamics from Superembedding}

\vspace{-0.25cm}


\subsection*{ Background}


Superembedding approach to supersymmetric extended objects provides a
natural and powerful geometrical framework for describing the dynamics
of superbranes. One of the most attractive aspects of this approach is
its universality; it seems to apply to any kind of branes, regardless of
whether their worldvolume multiplets are scalar, vector, tensor or,
indeed, any other supermultiplet. Another virtue of this approach is
that the target space {\it and} worldvolume supersymmetry are both made
manifest.

In the superembedding approach to branes, the brane under consideration is
described mathematically as a sub-supermanifold (the worldsurface) of the
target superspace. The coordinates transverse to the worldsurface are then the
Goldstone superfields which encode the information about the worldsurface
supermultiplets. The key point is then to impose a natural geometrical
constraint on the embedding which can translate into a constraint on the
Goldstone superfield. Indeed, a constraint of this nature does exist, and it
will be explained below.

The superembedding approach has its origin in what is known as the twistor-like
approach to superparticles/strings/branes. This approach was initiated some
time ago \cite{e1,e2} in the context of superparticles.   The formalism has
been extended to branes and it has been developed  by several authors over the
years. In particular we refer to \citelow{bsm2} and \citelow{hs1} for an
extensive list of references.

Starting with \citelow{hs1}, in a series of papers \cites{hs1}{lb}, the
superembedding formalism has been developed further. In particular, in
\citelow{hs1}, the nature of the worldsurface supermultiplets emerging from the
embedding formalism was spelled out. As applications, the full covariant field
equations of the M theory fivebranes were obtained for the first time by using
this formalism. Moreover, the existence of new types of superbranes were
deduced and/or conjectured within this framework. Later, this formalism was
used to describe open superbranes. The superembedding formalism yields
naturally the field equations rather than an action. However, it is possible
to obtain an action as well, and in a recent work the approach of \cite{} has
been extended to cover essentially any superbrane that does not involve
worldsurface self-dual field strengths. We will come back to these points
briefly later. First, let us begin with the description of the basics of
superembedding formalism, followed by some examples.

\vspace{-0.25cm}


\subsection*{ Basics of the Superembedding Formalism }


We consider superembeddings $f:M\rightarrow \unM$, where the
worldsurface $M$ has (even$|$odd) dimension $(d|{1\over2}D')$ and the
target space has dimension $(D|D')$. In local coordinates $M$ is given
as $z^{\unM} (z^M)$, where $z^{\unM}=(x^{\unm},\th^{\um})$ and
$z^{M}=(x^{m},\th^{\m})$ (if no indices are used we shall distinguish
target space coordinates from worldsurface ones by underlining the
former). The embedding matrix $E_A{}^{\unA}$ is defined to be

\be
E_A{}^{\unA}= E_A{}^M \,\left(\del_M z^{\unM}\right)\, E_{\unM}{}^{\unA}\ ,
\label{1}
\ee

in other words, the embedding matrix is the differential of the
embedding map referred to standard bases on both spaces. Our index
conventions are as follows: latin (greek) indices are even (odd) while
capital indices run over both types; letters from the beginning of the
alphabet are used to refer to a preferred basis while letters from the
middle of the alphabet refer to a coordinate basis, the two types of
basis being related to each other by means of the vielbein matrix
$E_M{}^A$ and its inverse $E_A{}^M$; exactly the same conventions are
used for the target space and the worldsurface with the difference that
the target space indices are underlined. Primed indices are used to
denote directions normal to the worldsurface. We shall also use a
two-step notation for worldsurface spinor indices where appropriate: in
general discussions, a worldsurface spinor index such as $\a$ runs from
1 to ${1\over2}D'$, but it may often be the case that the group acting
on this index includes an internal factor as well as the spin group of
the worldsurface; in this case we replace the single index $\a$ with the
pair $\a i$ where $i$ refers to the internal symmetry group. A similar
convention is used for normal spinor indices.

\fd

The basic embedding condition is

\be
E_{\a}{}^{\una}=0\ .
\label{2}
\ee

It implies that the odd tangent space of the worldsurface is a subspace
of the odd tangent space to $\unM$ at each point in $M\subset \unM$. In
many cases, equation \eq{2} determines the equations of motion for the
brane under consideration. Moreover, it also determines the geometry
induced on the worldsurface and implies constraints on the background
geometry which arise as integrability conditions for the existence of
such superembeddings. The fact that all this information can be deduced
from the simple equation \eq{2} can be intuitively be understood by observing
that the curl of the embedding matrix \eq{1} yields the formula
\be
\nabla_A E_B{}^{\unC}-(-1)^{AB}\nabla_{B} E_{A}{}^{\unC} +T_{AB}{}^C
E_C{}^{\unC}
=(-1)^{A(B+\unB)}E_B{}^{\unB} E_A{}^{\unA} T_{\unA\unB}{}^{\unC}
\label{de}
\ee

which involves the super torsion tensors of both, the world and target superspaces.
Thus, it is clear that feeding the basic embedding constraint \eq{2} into this
integrability equation will have consequences for  the worldvolume and target space
supergeometries, and hence the dynamics.

Remarkably, the basic embedding equation \eq{2} turns out to be sufficient to
determine the full covariant equations of motion for the collective coordinates
of the superbrane for most cases. For example, this is the case for both the
$M2$ and $M5$ branes. A qualitative aspect of this is that the larger the
codimension of the embedding is the stronger is the constraint \eq{2}. In fact,
it was found in \citelow{hs1} that three types of multiplet can arise as a
consequence of \eq{2}: on-shell, off-shell or underconstrained. In the on-shell
case, there can be no superspace actions of the heterotic string type since
such actions would necessarily involve the propagation of the Lagrange
multipliers that are used in this construction. Nevertheless, on-shell
embeddings are useful for deriving equations of motion; for example, the full
equations of motion of the $M$-theory fivebrane were first obtained this way
\cite{hs2}. In the off-shell case, by which it is meant that the worldsurface
multiplet is a recognisable off-shell multiplet, it is possible to write down
actions of the heterotic string type. The third case that arises, and which we
call underconstrained here, typically occurs for branes with low codimension.
For example, in codimension one the basic embedding condition gives rise to an
unconstrained scalar superfield. In order to get a recognisable multiplet
further constraints must be imposed. An example of this is given by IIA
D-branes where the basic embedding condition yields an on-shell multiplet for
$p=0,2,4$, but an underconstrained one for $p=6,8$. Imposing a further
constraint which states that there is a worldsurface vector field with the
usual modified Bianchi identity whose superspace field strength vanishes unless
all indices are bosonic, one recovers on-shell vector multiplets \cite{db}.
(For $p=0,2,4$ one can show that the vector Bianchi identity follows from the
basic embedding condition.)

The description of the superembedding formalism given above may understandably
give the impression that it provides only the superbrane field equations but
not an action from which they can be derived. In fact, a rather universal
method  has been proposed recently \cite{hrs2} by which a superspace action can
be obtained for a large class of superbranes. See \citelow{hrs2} for a detailed
description of the method and comparison with the work of \citelow{a1}.

The power of superembedding formalism has also been put to use in (a) the
derivation of the M$5$--brane equations of motion from those of an open
M$2$--brane ending on the M$5$--brane \cite{cs} and the dynamics of
$Dp$--branes ending on D$(p+2)$--branes \cite{chs,chsw}, and (b) the derivation
of a Born-Infeld type action for branes involving a higher than 2-form
potentials in their worldvolumes \cite{lb}. These branes have been called the
L-branes \cite{hs1} because their worldvolume fields form supermultiplets
known as the linear multiplets. For example, the L$5$--brane in $D=9$ has the
linear multiplet on  the worldvolume which contains a 4-form potential, and
action has been obtained for this object in \citelow{lb}.

\vspace{-0.25cm}


\subsection*{ Deriving the Field Equations from the Superembedding
Constraint }


In order to get a feel for how the embedding condition \eq{2} really determines
the worldvolume supermultiplet field equations, it suffices to study the
linearised version of the constraints resulting from the embedding condition in
flat target space limit.  The supervielbein for the flat target superspace is,

\bea
E^{\una}&=&dx ^{\una} -{i\over2} d\th^{\ua}(\C^{\una})_{\ua\ub}\th^{\ub}
\nn\\
E^{\ua}&=& d\th^{\ua}\ .
\eea

Let us choose the physical gauge,

\bea
x^{\una}&=&\cases{x^a &\cr x^{a'}(x,\th)  &\cr}
\nn\\
&&\nn\\
\th^{\ua}&=&\cases{\th^{\a}&\cr\th^{\a'}(x,\th)&\cr}
\eea

and take the embedding to be infinitesimal so that $E_A{}^M\del_M$ can
be replaced by $D_A=(\del_a,D_{\a})$ where $D_{\a}$ is the flat
superspace covariant derivative on the worldsurface, provided that the
embedding constraint holds. In this limit the embedding matrix is:

\be
\ba{rclrcl}
E_a{}^{\unb} &=& (\d_a{}^b ,\,\del_a X^{b'})\ ,&\quad\quad
E_a{}^{\ub}&=& (0,\,\del_a\th^{\b'})
\\
&&&&&\\
E_{\a}{}^{\unb} &=& (0,\, D_{\a}X^{a'}-i(\C^{a'})_{\a\b'}\th^{\b'})\ ,
&\quad\quad
E_{\a}{}^{\ub}&=& (\d_{\a}{}^{\b},\, D_{\a}\th^{\b'})\ , \la{le}
\ea
\la{lin}
\ee

where

\be
X^{a'}:=x^{a'}+{i\over2}\th^{\a}(\C^{a'})_{\a\b'}\th^{\b'}\ .
\ee

Using the expressions given in \eq{le} in the embedding condition \eq{2}, we
find, at the linearized level,

\be
D_{\a}X^{a'}=i(\C^{a'})_{\a\b'}\th^{\b'}\ . \la{mc}
\ee

This is a general formula. Next, we consider the examples of $M2$ and
$M5$ branes. Interestingly, the same embedding constraint \eq{mc} yields
the on-shell field equations of $N=8$ worldvolume scalar supermultiplet
in the M2 brane case, consisting of $8$ Bose and $8$ Fermi on-shell
physical degrees of freedom, and the $(2,0)$ worldvolume tensor multiplet
in the case of $M5$ brane, consisting of $5$ real scalars, a two-form
potential with self-dual field strength describing $3$ on-shell degrees
of freedom and $8$ fermionic on-shell degrees of freedom. Let us show
how this works, starting with the case of $M2$ brane.

\vspace{-0.25cm}

\subsection*{The M2 Brane}

The $M2$ brane worldvolume is an $(3|16)$ dimensional supermanifold
embedded in the target superspace of dimension $(11|32)$. The index
$\a=1,...,16$ which labels worldsurface fermionic coordinates carry a
spinor representation of $SO(2,1)\times SO(8)$, which will be denoted by
a pair of indices $\a A$ where $\a=1,2$ labels two component Majorana
spinor of $SO(2,1)$ and $A=1,...,8$ labels the chiral spinor of $SO(8)$.
The index $\a'=1,...,16$ labels the fermionic directions that are
normal to the worldsurface which will be denoted by $\a \dA$, where
$\dA=1,...,8$ labels the anti-chiral representation of $SO(8)$. The
master constraint \eq{2} can then be written as

\be
D_{\a A} X^{a'} = i (\s^{a'})_{A\dB}\,\th_\a^{\dB}\ ,
\ee

where $\s^{a'}$ are the chirally projected Dirac $\C$-matrices of
$SO(8)$ (the van der Wardeen symbols). Differentiating both sides with
$D_{\b B}$ and using the algebra of supercovariant derivatives

\be
\{ D_{\a}, D_{\b} \} = i \, (\c^a)_{\a\b} \del_a\ , \quad\quad a=0,1,2\ ,
\la{da1}
\ee

after straightforward manipulations one finds the result

\be
D_{\a A} \th_\b^{\dC}
= (\s^{a'})_A{}^{\dC}\,(\c^a)_{\a\b}\,\del_a X^{a'}\ . \la{mc2}
\ee

The equations of motion now arise as follows. Differentiating \eq{mc2}
with $D_{\c B}$, using \eq{mc}, symmetrizing the equation in $\c\a$
indices, using the algebra \eq{da1} of supercovariant derivatives and
multiplying with the $SO(2,1)$ charge conjugation matrix $\e^{\a\b}$, we
obtain the Dirac equation

\be
(\c^a)_\b{}^\c\,\del_a\th_\c^{\dC} = 0\ .  \la{dirac1}
\ee

Acting on \eq{mc2} with $(\c^b)^{\a\b} \del_b$, and using the Dirac
equation \eq{dirac1}, on the other hand gives the Klein-Gordon equation

\be
\del^a\del_a\,X^{a'}=0\ . \la{kg1}
\ee

Continuing in this manner, it can be shows that no new components arise
in the Goldstone superfields $X^{a'}$ and $\th_\b^{\dC}$. Thus, what we
have found is an $N=8$ on-shell scalar supermultiplet with $8$ real
scalars obeying the Klein-Gordon equation and $8$ two-component Majorana
spinors obeying the Dirac equation, altogether representing the $8$ fermi
and $8$ bose on-shell degrees of freedom on the $M2$ brane worldvolume.

\vspace{-0.25cm}

\subsection*{The M5 Brane}

The procedure for analysing the constraint \eq{mc} for the case of $M5$
brane is parallel to the case of $M2$ brane just described in
detail. Here, the $M5$ brane worldvolume is an $(6|16)$
dimensional supermanifold embedded in the target superspace of dimension
$(11|32)$. The index $\a=1,...,16$ which labels worldsurface fermionic
coordinates carry a spinor representation of $SO(5,1)\times SO(5)$,
which will be denoted by $\th_{\a i}$ where $\a=1,...,4$ is the chirally
projected spinor index of $SO(5,1)$ and $i=1,...,4$ labels the spinor of
$SO(5)$. The index $\a'=1,...,16$ labels the fermionic coordinates that
are normal to the worldsurface which will be denoted by $\th_i^\a$. We
are using the well established chiral notation in which the lower $\a$
index denotes a chiral spinor, the upper $\a$ index denotes and
anti-chiral spinor, and these indices are never to be raised and lowered by
a charge conjugation matrix. Furthermore, all the spinors in question are
symplectic Majorana-Weyl. See \citelow{hs1} for further notation and
conventions.

To analyze the constraint \eq{mc} for the case of $M5$ brane, we begin
by writing it more explicitly as

\be
D_{\a i}\,X^{a'} = i\,(\c^{a'})_{ij} \th_\a^j\ ,\quad a'=1,...,5\ ,\quad
\a=1,...,4\ , \la{mc5}
\ee

where $(\c^{a'})_{ij} $ are the Dirac $\c$-matrices of $SO(5)$ (which
are antisymmetric). The raising and lowering of the $SO(5)$ spinor index
is with the antisymmetric charge conjugation matrix $\O_{ij}$.

Starting from \eq{mc5}, manipulations parallel to those
described above for the case of $M2$ brane now lead to the result \cite{hs1}

\be
D_{\a i}\,\th_{\b j} =  -\ft12 (\c^{a'})_{ij} (\c^a)_{\a\b} \del_a X_{a'}
+\O_{ij} h_{\a\b}\ , \la{dt2}
\ee

where $(\c)_{\a\b}$ are the chirally projected Dirac $\c$-matrices of
$SO(5,1)$, and the symmetric bispinor $h_{\a\b}$ defines a self-dual
third-rank antisymmetric tensor

\be
h_{\a\b} \equiv (\c^{abc})_{\a\b}\,h_{abc}\ ,  \qquad
h_{abc}= \ft1{3!} \e_{abcdef} h^{def}\ .
\ee

Comparing the result \eq{dt2} with \eq{mc2}, reveals that the difference
between the $M2$ and $M5$ branes is due to the occurrence of a new
worldvolume field $h_{abc}$ in the latter case. Continuing in the manner
described for the $M2$ brane case earlier, one finds by applying further
spinorial covariant derivatives that the fermion field satisfies the
Dirac equation, the scalar fields satisfy the Klein-Gordon equation and
the tensor field satisfies the Bianchi identity and field equation for a
third-rank antisymmetric field strength tensor. Furthermore, there are
no other spacetime components, so that equation (\ref{mc}) defines an
on-shell tensor multiplet. It is remarkable that this result follows
from the superembedding constraint which takes exactly the same form for
both the $M2$ brane as well as the $M5$ brane. This shows the universal
nature of the embedding approach; although the worldvolume
supermultiplets are rather different in nature, they both arise from one
universal superembedding constraint. Unlike the Green-Schwarz type
formulation of branes in which one has to search for different kinds of
actions depending on the nature of the expected worldvolume
supermultiplet, here one starts from a universal and geometrical
embedding formula which then determines the worldvolume supermultiplet
and provides their equations of motion, if the codimension of the
embedding is large enough to make the constraint sufficiently strong to
do so. We just saw that this is the case for the $M2$ and $M5$ branes
which are described by codimension $8$ and $5$ embeddings.

\vspace{-0.25cm}

\subsection*{The L5 Brane}

As a last example to illustrate the universality of the superembedding
approach, we examine the $L5$ brane in $D=9$ \cite{hs1,lb}. The
$(6|8)$ dimensional worldvolume superspace is embedded in $(9|16)$
dimensional target superspace. This is a codimension 3 embedding in which
the 3 Goldstone superfields give rise to a linear supermultiplet with
$(1,0)$ supersymmetry in the $L5$ brane worldvolume. This multiplet
consists of $3$ real scalars, a $4$-form potential describing $5$
degrees of freedom and an $Sp(1)$ symplectic Majorana spinor describing
$8$ real degrees of freedom. This is an example for a superembedding in
which the embedding constraint is not sufficient to put the theory of-shell. As
a consequence, it is easier to write done an action formula for this theory.

The calculations at the linearized level are again very similar to those
explained in detail for the case of $M2$ brane earlier, so it suffices
to outline briefly how the worldvolume supermultiplet arises.

The index $\a=1,...,8$ now labels worldsurface fermionic coordinates which
carry a spinor representation of $SO(5,1)\times SO(3)$, which will be denoted
by $\th_{\a i}$ where $\a=1,...,4$ is the chirally projected spinor index of
$SO(5,1)$ and $i=1,2$ labels the spinor of $SO(3)$. The index $\a'=1,...,8$
labels the fermionic coordinates that are normal to the worldsurface which will
be denoted by $\th_i^\a$. The chiral notation for the spinors is as explained
earlier for the case of $M5$ brane. Thus, the master embedding constraint
\eq{2} again takes the form  \eq{mc5}. The only difference is that the index
$i=1,2$ now labels  an $SO(3)$ spinor. Steps parallel to those described above
then lead to the formula

\be
D_{\a i} \th_{\b j} = -(\c^{a'})_{ij} (\c^a)_{\a\b} \del_a X^{a'}
                         + \e_{ij} (\c^a )_{\a \b}\,h_a \ ,
\label{24}
\ee

where $h_a$ is the conserved vector in the multiplet, $\del^a h_a=0$. This
field, together with the 3 scalars $X^{a'}$ and the 8 spinors $\Th_{\a i}$
(evaluated at $\th=0$) are the components of the (off-shell) linear multiplet.
At the linearized level the field equations are obtained by imposing the free
Dirac equation on the spinor field. One then finds the Klein-Gordon equation
$\del_a \del^aX_{ij}=0$ for the scalars and the field equation for the
antisymmetric tensor gauge field $\del_{[a} h_{b]}=0$. The full  equations of
motion can be obtained \cite{lb} either by directly imposing an additional
constraint in superspace or by using the recently proposed brane action
principle which has the advantage of generating the modified Born-Infeld term
for the tensor gauge fields in a systematic way \cite{hrs2}.


\section{Supermembranes in AdS Background, Singletons and Higher
Spin Gauge Theory}


String theory has often been studied in Minkowski target spacetime or in a
product of Minkowski spacetime with tori, orbifolds or Ricci flat spaces such
as K3 or Calabi-Yau manifolds. These are spaces which allow a perturbative
formulation of string theory to all orders in $\alpha'$ as a conformal field
theory on the string worldsheet. Group manifolds also allow an exact conformal
field theoretic treatment and they have been studied in the context of string
theory as well, though to somewhat lesser extent. An additional motivation for
focusing attention on Ricci flat spaces has been the fact that
phenomenologically the most promising string theory is the heterotic string
theory which has natural compactifications that require Ricci-flat internal
spaces.

The study of duality symmetries in the early 90's and  the discovery of
D-branes in 1995 brought the type II theories under focus. The most often
studied type II backgrounds continued to be Minkowski $\times$ flat or Ricci
flat spaces for sometime but that changed drastically with the discovery in
1997 of a remarkable connection between type IIB string on $AdS_5\times S^5$
and $D=4,N=4$ supersymmetric $SU(N)$ Yang-Mills theory \cite{malda}. The AdS
background has emerged as the near horizon geometry of certain brane solutions,
and connections with Yang-Mills have been found by taking particular limits in
the parameter space of the theory. The study of branes in AdS space is now in
full swing, and it brings together nicely many aspects of brane physics,
supersymmetric field theories, Kaluza-Klein supergravities, gauged and
conformal supergravities. It has the further dividend of giving new handles on
old problems in nonperturbative Yang-Mills gauge theory.

The study of branes in AdS is not altogether a new development though. Already
back in late 80's, the $D=11$ supermembrane was studied in AdS background.
Indeed, the solutions of the $D=11$ supermembrane equations were studied in a
series of papers \cite{mc1,mc2,dps} in $M_4 \times M^7$ background, where $M_4$
was taken to be $AdS_4$ or its suitable covering and $M_7$ to be a suitable
seven dimensional Einstein space, such as $S^7$. Particularly interesting
solution was found in which a static spherical membrane resided at the boundary
of $AdS_4$. This was named the Membrane at the End of the Universe. As
mentioned earlier, the hope was that a perturbative expansion of the
supermembrane around this solution  would give a free field theory at the
boundary of AdS, thereby having significant consequences for the
renormalizability issue. At the time, the properties the full supermembrane
action in AdS background (that is, without expanding around a particular
solution) were not investigated. Recent developments, however, have provided
abundant motivation to do just that. It is convenient to discuss some general
features of the supermembrane in AdS space before we turn to a description of
the Membrane at the End of the Universe.

\vspace{-0.25cm}


\subsection*{Supermembrane in AdS Background and Singletons}


For definiteness, let us consider the supermembrane in $AdS_4 \times S^7$
background, which is a well known $N=8$ supersymmetric solution of $D=11$
supergravity \cite{dnp}. The $D=11$ supermembrane action in a generic
background is given by \cite{bst1}

\be
S=-\int d^3 \xi \left ( \sqrt{-g} + \e^{ijk} C_{ijk}\right)\ ,\la{action}
\ee

where $\x^i~(i=0,1,2)$ are the coordinates on the membrane worldvolume,
$g_{ij}$ is the induced metric on $\S$ and $g={\rm det}\, g_{ij}$. This metric
and the components of the pulled-back 3-form $C$ are defined as

\be
g_{ij} =E_i{}^{\una} E_j{}^{\unb} \eta_{\una\unb}\ ,
\quad\quad
C_{ijk} = E_i{}^{\unA} E_j{}^{\unB} E_k{}^{\unC} C_{\unC\unB\unA}\ ,
\la{cbg}
\ee

where $\eta_{\una\unb}$ is the Minkowski metric in eleven dimensions, and

\be
E_i{}^{\unA}= \del_i z^{\unM} E_{\unM}{}^{\unA}\ , \la{ee}
\ee
and $E_{\unM}{}^{\unA}$ is the target space supervielbein.

Thus, the $OSp(8|4)$ invariant supermembrane action  is \eq{cbg} in a target
superspace with isometry  group $OSp(8|4)$ which supports a closed 4-form
$dH=0$, which can be locally solved as $H=dC$. The superspace we seek must have
$AdS_4 \times S^7$ as a bosonic subspace and consequently it can be chosen
to be \cite{phu,rk3}

\be
{G\over H} = {OSp(8|4)\over SO(3,1) \times SO(7)}\ .\ \la{coset}
\ee

The generators of $G$ and $H$ are

\bea
G:\quad\quad && \overbrace{M_{ab}, P_c}^{SO(3,2)}\ ,\
               \overbrace{T_{IJ}, P_J}^{SO(8)}\ ,\  Q_{\a A} \nn
\w3
H:\quad\quad && M_{ab}\ , \ \ T_{IJ}
\eea

where $Q_{\a A}$ are the 32 real supergenerators transforming as spinor of
$SO(3,2)\times SO(8)$ and the rest of the notation is self explanatory. The
supervielbein and the 3-form $C$ on $G/H$ can be calculated straightforwardly
from the knowledge of the structure constants of $G$. See
\citelow{rk3,fre,dewit} for further details.

The action \eq{action}, with target superspace $G/H$ specified in \eq{coset},
is manifestly invariant under $OSp(8|4)$ since this is the isometry group of
$G/H$. It is also invariant under the worldvolume local diffeomorphism and
local $\k$--symmetry. Fixing a physical gauge by identifying the worldvolume
coordinates with three of the target space coordinates and setting half of the
target space fermionic coordinates  equal to zero (by means of a suitable
projection), breaks the local diffeomorphisms, local $\k$--symmetry as well as
the rigid isometries of $G/H$. The requirement of maintaining the physical gauge
fixes the local symmetry  parameters in terms of the rigid parameters, and
consequently one arrives at a gauged fixed worldvolume action which is
invariant under the rigid $G$ symmetry \cite{rk2,fre}.

Thus one obtains an action for the $8$ real scalars and $8$  Majorana spinors
on the worldvolume, which is invariant under the rigid superconformal group
$OSp(8|4)$ transformations some of which are linearly realized (and hence
manifest) and the rest are nonlinearly realized. All this is perfectly
analogous to the discussion of the lightcone gauge fixing in $D=11$
supermembrane theory in Poincar\'e superspace \cite{bst2}.

Now, we ask the following question: Is there a vacuum solution of the
supermembrane equation such that a perturbative expansion around it yields a
free but still $OSp(8|4)$ invariant action? The answer is yes. To see this, it is
convenient to use horospherical $\times$ hyperspherical coordinates
to parametrize the $AdS_4\times S^7$ metric as

\be
ds^2=\phi^2\,\left(-d\tau^2+d\s^2+d\rho^2\,\right) + a^{-2}
\,\left({d\phi \over \phi}\right)^2  + 4a^{-2}\, d\Omega_7\ ,
\ee

where $d\Omega_7$ is the $SO(8)$ invariant metric on $S^7$. The boundary of
this metric is the three dimensional Minkowski space $M_3$ at $\phi\rightarrow
\infty$, completed by the point $\phi\rightarrow 0$, so that the inversion
element of the conformal group action on the boundary is well defined. The
nature of this boundary has been discussed in great detail, for example, in
\citelow{ew,rk2,fre}.

Denoting the coordinates of the seven sphere by $y^I$ (I=1,...,7), a simple
class of solutions to the supermembrane equations is \cite{fre}

\be
x^i=\xi^i\ , \quad\quad \phi=\phi_0\ ,\quad\quad  y^I=y^I_0\ ,
\la{sol2}
\ee

where $x^i=(\tau,\sigma,\rho)$ and $\xi$ are the worldvolume coordinates,
$\phi_0$ and $y^I_0$ are arbitrary constants and the fermionic variables are set
equal to zero. The singleton action is obtained by expanding around $\phi_0=0$
which corresponds to expanding around the boundary of $AdS_4$ \cite{rk2,fre}. For
convenience, one can also set $y_0^I=0$. In using the normal coordinate expansion
formulae \cite{ss13}, it is important to rescale the fluctuation fields
appropriately. Defining the fluctuations as \cite{fre}

\be
\phi=\phi_0 +  \sqrt {\phi_0}\,\varphi\ , \qquad y^I= {1\over
\sqrt{\phi_0}}\,\varphi^I\ , \qquad \th = (\phi_0)^{-3/2}\,\lambda\ ,
\ee

and taking the limit $\phi_0 \rightarrow 0$ after substituting to the  normal
coordinate expansion formulae \cite{ss13}, one finds that the zeroth and first
order terms in the normal coordinate expansion vanish and the second order
term  yields the $N=8$ singleton action for eight free scalars
$(\varphi,\varphi^I)$ and eight free fermions $\lambda$ \cite{rk2,fre}. The
action is $OSp(8|4)$ invariant and this does not require  a mass term  for the
bosons, since the boundary of $AdS_4$ is characterized as a  Minkowski space in
the coordinate system used here.

\vspace{-0.25cm}


\subsection*{Membrane at the End of the Universe}


Now, we turn to the description of an interesting set of solutions to the
supermembrane equations in $AdS_4 \times S^7$ background which were found
sometime ago \cite{mc1,mc2} and which are closely related to solution
\eq{sol2}. We begin by considering the bosonic field equation for a
configuration where the spacetime gravitino and the fermionic co-ordinates are
set equal to zero:

\bea
&&\partial_i(\sqrt{-h}h^{ij}\partial_j X^Ng_{MN}) - {1\over 2}\sqrt{-h}
h^{ij}\partial_iX^N\partial_j X^P \partial_M g_{NP}\cr
&&+{1\over 3}\epsilon^{ijk}\partial_i
X^N\partial_j X^P\partial_k X^QH_{MNPQ}=0\ , \la{mfe}
\eea

where $h_{ij}$ is the induced metric

\be
h_{ij}=\del_i X^M \del \del_j X^N G_{MN}\ ,
\ee

$X^M(\tau,\sigma,\rho)$ are the spacetime coordinates of the membrane, $g_{MN}$
is
the spacetime metric and  $H_{MNPQ}=4\del_{[M}B_{NPQ]}$. The $\k$--symmetry of
the
supermembrane action  requires that $g_{MN}$ and $H_{MNPQ}$ satisfy the usual
bosonic equations of $d=11$ supergravity. Let us consider the solution of these
equations is which the spacetime is $AdS_4\times M_7$, such that $AdS_4$ has
inverse
radius $a$, and $M_7$ is an Einstein space with Ricci tensor $R_{mn}={3\over 2}
a^2
g_{mn}$. In \citelow{mc2} we used the coordinate system in which the $AdS_4$
metric
is given by

\be
ds^2=-(1+a^2r^2)dt^2+r^2(d\theta^2+\sin^2\theta d\phi^2)
+(1+a^2r^2)^{-1}dr^2\ .
\ee

The maximal $SO(3,2)$ symmetry of this metric requires that the period of
$t$ is an integer multiple of $2\pi/a$, namely $\Delta t=2\pi q/a\ ,
\la{defq}$ where $q$ is an integer \cite{mc2}. The four-index field strength
is taken to be proportional to the Levi-Civita tensor in $AdS_4$:

\be
H_{\m\n\r\s}=\ft32\,a \sqrt {-g}\ \e_{\m\n\r\s}\ .
\ee

The metric for the seven dimensional internal space will be taken to be an
$S^7$ moded by $Z^p$, which is a Lens space, $L(1,p)$ that can also be
viewed as a $U(1)$ bundle over $CP(3)$ with fibers having the period
$2\pi/p$, for some integer $p$:

\be
d\hat s^2 = {4\over{a^2}}[(d\psi+2A)^2+ds^2(CP^3)]\ , \la{m1}
\ee

where $ds^2(CP^3)$ is the standard Fubini-Study metric on $CP^3$, with
Einstein metric satisfying $R_{ab}=8g_{ab}$, $\psi$ is the coordinate on the
U(1) fibres and $A$ is a one-form potential satisfying $dA = J$, where $J$
is the K\"ahler two-form on $CP^3$. The fibre coordinate $\psi$ has the
period $ \Delta \psi=2\pi/p$, for some integer $p$. The space $L(1,1)$ is
the round seven sphere $S^7$, and $L(1,2)$ is the projective space $RP^7$.

To solve the supermembrane equations, we make the ansatze \cite{mc2}

\bea
t &=&\tau\ , \quad\quad\ \ \th =\sigma\ , \quad\quad\qquad \phi= \rho\ ,
\nn\\
r &=& r_0 \ ,\quad\quad  y^I= y_0^I\ , \quad\quad\quad\  \psi = \a\tau/2\ ,
\la{ca}
\eea

where $\alpha, r_0$ and $y_0^I$ are constants, and $y^I$ are the coordinates on
$CP^3$, when we view the internal space as $U(1)$ bundles over $CP^3$.
Substituting this ansatze into the supermembrane field equations \eq{mfe},  we
find that all components of the equation are satisfied identically except in
the $r$ direction, which is solved by $\ (\,\alpha=1,\ r_0=arbitrary\,)$, or
by  $\ (\,\a >1, \ r_0 =\sqrt {4(\a^2-1)/3a^2}\,)$. Furthermore we note that
the periods of $\t$ and $\psi$ specified above, together with the ansatz
\eq{ca}, imply that $\a=2/pq$. Thus we find the following solutions \cite{mc2}

\bea
&&\a=1:\qquad\quad  \widetilde {AdS_4}\times S^7\ ,
\quad\qquad r_0= {\rm arbitrary}\ ,\nn
\w2
&&\quad\qquad\qquad\quad\  AdS_4 \times RP^7\ ,\qquad r_0= {\rm arbitrary}\ ,\nn
\w2
&&\a=2:\qquad\ \  AdS_4 \times S^7 \ ,\quad\qquad r_0=2/a \ , \la{sols}
\eea

where $\wt{AdS_4}$ denotes the double covering of $AdS_4$. As was shown in
\citelow{mc2}, the supersymmetry of the solutions presented above requires that
$r_0$ be taken to infinity (hence the terminology of the ``Membrane at the End
of the Universe''). This requirement peaks the $\a=1$ solutions above. A normal
coordinate expansion of the action around these solutions should yield an
$OSp(8|4)$ invariant action formulated on the boundary of $AdS_4$ (or
$\wt{AdS_4})$. However, it has been noted \cite{mc2} that the
rescalings  of the fluctuation fields needed to extract a nonvanishing
quadratic action (which also eliminate all the interaction terms in the limit
$r_0 \rightarrow \infty$) had the effect of spoiling the extraction of meaningful
supersymmetry transformations (which required different rescalings). Thus, a
rigorous derivation of the action is still missing. Nonetheless, we expect the
standard $OSp(8|4)$ singleton field theory on the boundary of $AdS_4$ to arise.
The action for this theory \cite{bd1,nst} can be described as follows.

The  $OSp(8|4)$ singleton supermultiplet consists of $8$ real scalars $\phi^I
(I=1,...,8)$ in the $8_v$ representation of $SO(8)$ and $8$ four-component
spinors $\l_-^\a\,(\a=1,...,8)$ in $8_s$ of $SO(8)$. These fields live on
$S_2\times S^1$ boundary of $AdS_4$. In addition to the four-dimensional
Majorana condition ${\bar \l}=\l^TC$, the spinor $\l$ satisfies the chirality
condition

\be
\c_0\c_1\c_2\,\l_-= -\l_-\ ,
\ee

which, unlike the usual chirality condition, is compatible with the Majorana
condition. The $N=8$ supersingleton action is given by \cite{bd1,nst}

\be
{\cal L}= -\ft12 {\sqrt -h} \left(\,h^{ij} \del_i \phi^I \del_j \phi^I +
\ft14\,a^2\,\phi^I \phi^I - i{\bar \l_-}^{\a} \c^i \nabla_i \l_-^{\a}
\, \right) \ , \la{n8}
\ee

where $\nabla_i$ is the covariant derivative on $S_2 \times S^1$. This action
is invariant under the rigid $OSp(8|4)$ transformations, the details of which
can be found in \citelow{bd1,nst,ss13}. Notice the presence of the mass
term for the bosonic singletons in the action. In the case of $S^p \times S^1$
boundary of $AdS_{p+2}$, this term is given by \cite{st}

\be
{p-1\over 4p}\, R\,\phi^2\ ,
\ee

where $R= p(p-1) a^2$ is the curvature scalar of $S^p\times S^1$ and $a$ is the
inverse radius of $S^p$. In fact, the difficulty in obtaining the Lagrangian
\eq{n8} from the normal coordinate expansion of the supermembrane action on
$AdS_4 \times S^7$ around the Membrane at the End of the Universe solutions
lies precisely in getting this mass term right, for it seems to vanish if one
takes a naive  $r_0 \rightarrow \infty $ limit \cite{mc2,rk2,duffrev}.


\section*{Singletons and Higher Spin Gauge Theory}


The $N=8$ singleton field theory formulated on $S^2 \times S^1$ boundary of
$AdS_4$ was quantized sometime ago \cite{ss12,ss13}, with the hope that it
might play a role in the  quantization of the supermembrane on $AdS_4 \times
S^7$ background \cite{bd1,ss13}.  Furthermore, a remarkable group theoretical
property of the singleton representation which states that the direct product
of two singletons decomposes into infinitely many massless field with higher
spins \cite{ff1}, motivated Bergshoeff, Salam, Tanii and the author
\cite{ss12,ss13,bst00} to conjecture that the $D=11$ quantum supermembrane on
$AdS_4 \times S^7$ should give rise to a higher spin gauge theory which
contains the usual $N=8$ AdS supergravity as a subsector. The occurrence of the
infinitely many massless higher spin fields implies the existence of infinitely
many (local) gauge symmetries analogous to the Yang-Mills, general coordinate
and local supersymmetries associated with spin 1, 2 and 3/2, respectively.

Interestingly enough, in a related development Fradkin and Vasiliev
\cite{fv0,fv1} were in the course of developing a higher spin gauge
theory in its own right (see \citelow{fv0} for references to earlier work).
These authors succeeded in constructing interacting field theories for
higher spin fields. It was observed that the previous difficulties in
constructing higher spin theories can be bypassed by formulating the
theory in AdS space and to consider an infinite tower of gauge fields
controlled by various higher spin algebras based on certain infinite
dimensional extensions of super AdS algebras. In particular, the AdS
radius could not be taken to infinity since its positive powers occurred
in the higher spin interactions and therefore one could not take a naive
Poincar\'e limit.

In a series of papers Vasiliev pursued the program of constructing the $AdS$
higher spin gauge theory and simplified the construction considerably. In
\citelow{v4} the spin 0 and 1/2 fields were introduced to the system within the
framework of free differential algebras. The theory was furthermore cast into
an elegant geometrical form \cite{v10} by extending the higher spin algebra to
include new auxiliary spinorial variables (see \citelow{v11} for a review).

Applying the formalism of Vasiliev to a suitable higher spin algebra that
contains the maximally extended super AdS algebra $OSp(8|4)$, the resulting
spectrum of gauge and matter fields remarkably coincide with the massless
states resulting from the symmetric product of two $OSp(8|4$ supersingletons!
\cite{ss12,ss13,bst00}. In a recent paper \cite{pers}, the Vasiliev theory
of higher spin fields (which is applicable to a wide class of higher spin
superalgebras) was examined in the context of $N=8$ supersymmetry, and  the
precise manner in which the $N=8$ de Wit-Nicolai gauged supergravity \cite{dn2}
can be described within this framework was studied.

In the rest of this section, we will outline some kinematical aspects of the
higher spin gauge theory. To begin with, the $N=8$ singleton representation
of  the $D=4$  super AdS group $OSp(8|4)$ \cite{g1,ns}, which are also known
as Di's and Rac's, are
\bea
Rac &:& \qquad  D(\ft12,0)\otimes 8_{s}  \nn
\w2
Di &:& \qquad D(1,\ft12)\otimes 8_{c} \nn
\eea

where $D(E_{0},s)$ denotes an UIR of $SO(3,2)$ for which $E_{0}$ is the
minimal energy eigenvalue of the energy operator $M_{04}$, and $s$ is the
maximum eigenvalue of the spin operator $M_{12}$ in the lowest energy sector.
The decomposition of the symmetric tensor product $\left[(Rac\oplus
Di)\otimes (Rac\oplus Di)\right]_{S}$ under the $OSp(8|4)$ leads to the UIR's
of $OSp(8|4)$ given in Table 1 \cite{ss12,ss13}.

\tb

The integer $k=0,1,2,..$ can be considered as level number.  Levels $\k=0,1$
are somewhat special.  At level $k=0$, there is the familiar $N=8$ supergravity
multiplet consisting of $128$ Bose and $128$ Fermi on-shell degrees of
freedom.  At level $k=1$, there is a $256+256$ multiplet that has $2$ scalars
as the lowest member of the supermultiplet and a single spin $4$ field as the
highest member.  For levels $k>2$ the structure of all the supermultiplets is
the same, namely they start with a singlet spin $(2k-2)$ field and end with a
singlet $(2k+2)$ field, in $8$ steps of spin 1/2 increments.  The associated
$SO(8)$ irreps are: $(1,8,28,56,35+35,56,28,8,1)$.

Among the interesting and important properties of the spectrum shown in
Table 1 is that there are two distinct classes one of fields; those with
spin $s \ge 1$, which forms an infinite set, and few fields that have spin
$s\le 1/2 $.  These two classes of fields are separated with a vertical line
in Table 1.  The fields with spin $s\ge 1$ can be associated with generators
of an infinite dimensional algebra, called $shs^E(8|4)$, while the fields
with spin $s\le 1/2$, clearly cannot be associated with any generator.  Note
however the important fact that all fields shown in Table 1, including
those with spin $s \le 1/2$ are exactly those which arise in the symmetric
tensor product of two $OSp(8|4)$ singletons. Physical consistency of gauge
field theory based on $shs^E(8|4)$ requires that the complete particle
spectrum forms a unitary representation of the full, infinite dimensional
algebra $shs^E(8|4)$ \cite{kv1}.  The product of the $OSp(8|4)$ singletons
which give the field content shown in Table 1, indeed does form a unitary
representation of $shs^E(8|4)$.  Consequently, the matter fields (the left
hand side of the vertical line shown in Table 1) must be included, in
addition to the gauge fields (the right hand side of the vertical line) in a
sensible (consistent and unitary) formulation of higher spin $N=8$
supergravity theory.

Recently \cite{pers}, the $N=8$ higher spin supergravity theory based on \alg
was investigated in considerable detail, and the precise manner in which it
contains the $N=8$ de Wit-Nicolai gauged supergravity \cite{dn2} has been shown
at the linearized level.  In our opinion, this constitutes a positive step
towards the understanding of the M-theoretic origin of the massless higher spin
gauge theory.  To make further progress, one has to compare the interactions of
the spin $s\le 2$ fields in the higher spin theory with those of de Wit-Nicolai
gauged supergravity theory at the next order, namely the quadratic order in
fields, in the equations of motion.  It would be very interesting to find out
how the $E_7/SU(8)$ structure of the scalar fields \cite{cj2} will manifest
itself and to determine how the higher spin fields interact with the spin $s\le
2$ fields.  Ultimately, the $N=8$ higher spin supergravity should emerge from
the dynamics of the $N=8$ singleton field theory defined at the boundary of
$AdS_4$. In this context, we note that the OPE's of the stress energy tensor
in the $OSp(8|4)$ singleton theory have been studied \cite{ope}, but a great
deal of work remains to be done to shed more light on the issue of how to
extract information about the physics in the bulk of AdS.

In summary, it is worth emphasizing the following points about the $N=8$
higher spin supergravity whose properties have been outlined above: (a) the
existence of the theory is highly nontrivial, (b) the theory is based on an
infinite dimensional extension of the $D=4,N=8$ super AdS group $OSp(8|4)$,
(c) it fuses matter fields with the gauge fields in such a way that the full
spectrum of massless states are exactly those which arise from the two $N=8$
singleton states and (d) the theory contains the equations of motion of the
$D=4,N=8$ AdS supergravity as a subsector. This last property is very
significant in that it is in the spirit of discovering new structures that
build upon what we already know. Should this theory survive further scrutiny,
then the appropriate question to ask is not if this theory fits into the big
picture of M-theory, but rather how it will do so.

The massive Kaluza-Klein states coming from the $S^7$ compactification of
$D=11$ supergravity must also be taken into account in a suitable extension of
the $N=8$ higher spin supergravity. These states are expected to arise in the
product of three or more $N=8$ singletons.  An infinite set of new massive
states would appear in the spectrum as a byproduct. A higher spin theory
taking into account massive states is yet to be constructed.

An extreme point of view would be to imagine a pure gauge theory formulation of
M-theory which contains only the massless fields corresponding to an infinite
dimensional symmetry. All the phases of M-theory (the known ones and those yet
to be discovered) are then to emerge from the breaking of this master symmetry
in various ways.

One can imagine the construction of a higher spin gauge theory directly in
$D=11$ AdS space. However, the anticommutator of two supersymmetries
necessarily involves (tensorial) generators in addition to the AdS generators
in $D=11$. One can take the AdS group in $D=11$ to be the diagonal subgroup of
$OSp(1|32)\oplus OSp(1|32)$ \cite{vhvp,g5} and study its field theoretic
realizations. Apparently, no such realizations  are known at present
\cite{df1,df2}. Nonetheless, the singleton representation for this group has
been studied recently by G\"unaydin \cite{g5} who found that the product of two
such irreps do not contain the $D=11$ supergravity states, but a further
product of the resulting representation does contain the $D=11$ supergravity
fields {\it and} additional fields as well. This group theoretical result
suggests the construction of a $10D$ singleton field theory that lives on the
boundary of the $D=11$ AdS space.

It is clear that much remains to be discovered and that these are exciting
times in the quest for an understanding of the mysterious and magic membrane
theory.

\section*{Acknowledgements}

I wish to thank L. Baulieu, E. Bergshoeff, M. Cederwall, C. Chu, M.
Duff, E. Eyras, M. G\"unaydin, J. Lu, B. Nilsson and P. Sundell for
useful discussions, and F. Mansouri for drawing my attention to the
developments in early 70's pertaining to the relation between
generalized dual models and theories of extended objects. I also would
like to thank the Abdus Salam International Center for Theoretical
Physics, University of Groningen, Universit\'es Pierre et Marie Curie
(Paris VI) and Denis Diderot (Paris VII) and the Institute of
Theoretical Physics in G\"oteborg for hospitality. This research has
been supported in part by NSF Grant PHY-9722090.

\newpage


\section*{References}


\begin{thebibliography}{99}

\bm{ir} {\it Ideals and Realities : Selected Essays of Abdus Salam},
         eds. C.H. Lai and A. Kidwai (3rd edition, World Scientific, 1990).

\bm{ir1} Abdus Salam, {\it Physics and the excellences of the life it
         brings}, reprinted in [1].

\bm{ir2} Abdus Salam, {\it Gauge unification of fundamental forces},
         reprinted in [1].

\bm{ir3} Abdus Salam, {\it Islam and science}, reprinted in [1].

\bm{ir4} Abdus Salam, {\it Particle physics: will Britain kill its own
         creation?''}, reprinted in [1].


\bm{ss1} A. Salam and E. Sezgin, {\it Maximal extended supergravity theory in seven
    dimensions}, Phys. Lett. {\bf 118B} (1982) 359.

\bm{ss2}  A. Salam and E. Sezgin, {\it SO(4) gauging of N=2 supergravity in seven
    dimensions}, Phys. Lett. {\bf 126B} (1983) 295.

\bm{ss3} A. Salam and E. Sezgin, {\it Chiral compactification on
    Minkowski $\times S^2$ of $N=2$ Einstein-Maxwell supergravity}, Phys.
    Lett. {\bf 147B} (1984) 47.

\bm{ss4} S. Randjbar-Daemi, A. Salam and E. Sezgin and J. Strathdee,
    {\it An anomaly free model in six dimensions}, Phys. Lett. {\bf 151B}
    (1985) 351.

\bm{ss5} A. Salam and E. Sezgin, {\it d=8 supergravity}, Nucl. Phys.
    {\bf B258} (1985) 284.

\bm{ss6} A. Salam and E. Sezgin, {\it d=8 supergravity: matter
    couplings, gaugings and Minkowski space compactifications}, Phys. Lett.
    {\bf 154B} (1985) 37.

\bm{ss7} A. Salam and E. Sezgin, {\it Anomaly freedom in chiral
    supergravities}, Physica Scripta {\bf 32} (1985) 283.

\bm{ss8} E. Bergshoeff, S. Randjbar-Daemi, A. Salam, H. Sarmadi and E.
    Sezgin, {\it Locally supersymmetric sigma model with Wess-Zumino term in
    two dimensions and critical dimensions for strings}, Nucl. Phys. {\bf
    B269} (1986) 77.

\bm{ss9} E. Bergshoeff, A. Salam and E. Sezgin, {\it A supersymmetric
    $R^2$-action in six dimensions and torsion}, Phys. Lett. {\bf
    173B} (1986) 73.

\bm{ss10} E. Bergshoeff, T.W. Kephardt, A. Salam and E. Sezgin, {\it
    Global anomalies in six dimensions}, Mod. Phys. Lett. {\bf A1} (1986) 267.

\bm{ss11} E. Bergshoeff, A. Salam and E. Sezgin, {\it Supersymmetric
    $R^2$-actions, conformal invariance and the Lorentz Chern-Simons term in
    6 and 10 dimensions}, Nucl. Phys. {\bf B279} (1987) 659.

\bm{ss12} E. Bergshoeff, A. Salam, E. Sezgin and  Y. Tanii, {\it
    Singletons, higher spin massless states and the supermembrane}, Phys.
    Lett. {\bf 205B} (1988) 237.

\bm{ss13} E. Bergshoeff, A. Salam, E. Sezgin and Y. Tanii, {\it N=8
    supersingleton quantum field theory}, Nucl. Phys. {\bf B305} (1988)
    497.

\bm{dd} A. Salam and E. Sezgin, {\it Supergravities in Diverse
    Dimensions}, Vols. 1 \& 2 (World Scientific, 1989).


\bm{diracm} P.A.M. Dirac, {\it An extensible model of electrons}, Proc. Roy.
            Soc. {\bf 268A} (1962) 57.

\bm{js1} J.H. Schwarz, {\it Superstrings. The First Fifteen Years}, Vols. 1 \&
         2 (World Scientific, 1985).

\bm{n1} G. Domokos, {\it Wave packet realization of lightlike states},
        Commun. Math. Phys. {\bf 26} (1972) 15.

\bm{n2} F. G\"ursey and S. Orfanidis, {\it Extended hadrons, scaling variables
        and the Poincar\'e group}, Nuovo Cimento {\bf 11A} (1972)225.

\bm{n3} E. Del Giudice, P. di Vecchia, S. Fubini and R. Musto, {\it Lightcone
        physics and duality}, Nuovo Cimento {\bf 12A} (1972) 813.

\bm{n4} F. Mansouri, {\it Dual models with global SU(2,2) symmetry},
        Phys. Rev. {\bf D8} (1973) 1159.

\bm{n5} R.J. Rivers, {\it Conformal algebraic explanation for the
        failure of dual theories with satellites}, Phys. Rev. {\bf D9} (1974)
        2920.

\bm{n6} F. Mansouri, {\it Natural extensions of string theories}, in
        {\it Lewes String Theory Workshop}, July 1985 , eds. L. Clavelli and
        A. Halpern (World Scientific, 1986).

\bm{cjs} E. Cremmer, B. Julia and J. Scherk, {\it Supergravity in eleven
    dimensions}, Phys. Lett. {\bf B76} (1978) 409.

\bm{gs1} M.B. Green and J.H. Schwarz, {\it Covariant description of
    superstrings}, Phys. Lett. {\bf B136}(1984) 367.

\bm{bst346} E. Bergshoeff, E. Sezgin and P.K. Townsend, {\it Superstring
    actions in $D=3,4,6,10$ curved superspace}, Phys. Lett. {\bf B169}
    (1986) 191.

\bm{hp} J. Hughes and J. Polchinski, {\it Partially broken global supersymmetry
       and the superstring}, Nucl. Phys. {\bf B278} (1986) 147.

\bm{hlp} J. Hughes, J. Liu and J. Polchinski, {\it Supermembranes}, Phys.
    Lett. {\bf 180B} (1986)370.

\bm{bst1} E. Bergshoeff, E. Sezgin and P.K. Townsend, {\it Supermembranes and
         eleven-dimensional supergravity}, Phys. Lett. {\bf 189B} (1987) 75.

\bm{bst2} E. Bergshoeff, E. Sezgin and P.K. Townsend, {\it Properties of
    the eleven dimensional supermembrane theory}, Ann. Phys.
    {\bf 185} 330.

\bm{at} A. Ach\'ucarro, J.M. Evans, P.K. Townsend and D.L. Wiltshire, {\it
        Super p-branes}, Phys. Let. {\bf 198B} (1987) 441.

\bm{duff87} M.J. Duff, P.S. Howe, T. Inami and K.S. Stelle, {\it
    Superstrings in D=10 from supermembranes in D=11}, Phys. Lett. {\bf
    B191} (1987) 70.

\bm{bps87} I. Bars, C.N. Pope and E. Sezgin {\it Massless spectrum and
    critical dimension of the supermembrane}, Phys. Lett. {\bf B198} (1987)
    455.
\bm{duff88} M.J. Duff, T. Inami, C.N. Pope, E. Sezgin and K.S. Stelle, {\it
    Semiclassical quantization of the supermembrane}, Nucl. Phys. {\bf
    B297} (1988)515.

\bm{hoppe} B. de Wit, J. Hoppe and H. Nicolai, {\it On the quantum mechanics of
    supermembranes}, Nucl. Phys. {\bf B305} (1988) 545.

\bm{bst90} E. Bergshoeff, E. Sezgin, Y. Tanii and P.K. Townsend, {\it Super
    p-branes as gauge theories of volume preserving diffeomorphisms}, Ann.
    Phys. {\bf 199} (1990) 340.

\bm{ms}U. Marquard and M. Scholl, {\it Lorentz algebra and critical dimension for
       the supermembrane}, Phys. Lett. 227B (1989) 234.

\bm{bars}I. Bars, {\it First massive level and anomalies in the supermembrane},
         Nucl. Phys. B308 (1988) 462.

\bm{bp} I. Bars and C.N. Pope, {\it Anomalies in super p-branes}, Class. and
    Quantum Grav. {\bf 5} (1988) 1157.

\bm{ps} C.N. Pope and E. Sezgin, {\it A generalized Virasoro algebra for the
    type IIA superstring}, Phys. Lett. {\bf B233}(1989)313.

\bm{bs1} E. Bergshoeff and E. Sezgin, {\it New spacetime algebras and
    their Kac-Moody extension}, Phys. Lett. {\bf B232} (1989) 96.

\bm{bs2} E. Bergshoeff and E. Sezgin, {\it Super p-branes and new spacetime
          superalgebras}, Phys. Lett. {\bf B354} (1995) 256, hep-th/9504140.

\bm{es3} E. Sezgin, {\it The M-algebra}, Phys. Lett. {\bf B392} (1997) 323,
    hep-th/9609086.

\bm{pktsol} P.K. Townsend, {\it Supersymmetric extended solitons}, Phys.
    Lett. {\bf B202} (1988) 53.

\bm{dab} A. Dabholkar, G.W. Gibbons, J.A. Harvey and F. Ruiz Ruiz, Nucl. Phys.
    {\bf B340} (1990) 33.

\bm{str1} A. Strominger, {\it Heterotic solitons}, Nucl. Phys. {\bf
    B343} (1990) 167.

\bm{ds} M.J. Duff and K.S. Stelle, {\it Multimembrane solutions of D=11
        supergravity}, Phys. Lett. {\bf B253} (1991) 113.

\bm{g} R. G\"uven, {\it Black $p$-brane solutions of D=11 supergravity
       theory}, Phys. Lett. {\bf B276} (1992) 49.

\bm{dl1} M.J. Duff and J.X. Lu, {\sl Type II p-branes: the brane scan
    revisited}, Nucl. Phys. {\bf B390} (1993) 276, hep-th/9207060.

\bm{dgt} M.J. Duff, G.W. Gibbons and P.K. Townsend, {\it Macroscopic
    superstrings as interpolating solitons}, Phys. Lett. {\bf B332} (1994) 321,
    hep-th/9505124.

\bm{sw1} N. Seiberg and E. Witten, {\it Monopole condensation, and
    confinement in $N=2$ supersymmetric Yang-Mills theory}, Nucl. Phys.
    {\bf B426} (1994) 19, hep-th/9407087.

\bm{hlw} P.S. Howe, N.D. Lambert and P.C. West, {\it Classical
    M-fivebrane dynamics and quantum $N=2$ Yang-Mills}, hep-th/9710034.


\bm{ht1} C. M. Hull and P. K. Townsend, {\it Unity of superstring
    dualities}, Nucl. Phys. {\bf B438} (1995) 109, hep-th/9410167.

\bm{pkt1} P.K. Townsend, {\it The eleven dimensional supermembrane
    revisited}, Phys. Lett. {\bf B350} (1995) 184, hep-th/9501068.

\bm{ew95} E. Witten, {\it String theory dynamics in various
    dimensions}, Nucl. Phys. {\bf B443} (1995) 85, hep-th/9503124.

\bm{pktdem} P.K. Townsend, {\it p-brane democracy}, hep-th/9507048.

\bm{pol} J. Polchinski, {\it Dirichlet-branes and Ramond-Ramond charges},
    Phys. Rev. Lett. {\bf 75} (1995) 4724, hep-th/9510017.

\bm{jhs0} J.H. Schwarz, {\it The power of M theory}, Phys. Lett. {\bf B367}
    (1996) 97, hep-th/9510086.

\bm{hw} P. Horava and E. Witten, {\it Heterotic and type I string
    dynamics from eleven dimensions}, Nucl. Phys. {\bf B460} (1996) 506,
    hep-th/9510209.

 \bm{ch1} C.M. Hull, {\it String dynamics at strong coupling}, Nucl.
    Phys. {\bf B468} (1996) 113, hep-th/9512181.

\bm{ah1} O. Aharony, J. Sonnenschein and S. Yankielowicz, {\it
    Interactions of strings and D-branes from M theory}, Nucl. Phys. {\bf
    B474} (1996) 309, hep-th/9603009.

\bm{mm} T. Banks, W. Fischler, S.H. Shenker and  L. Susskind, {it M theory as a
    matrix model: A conjecture}, Phys. Rev. {\bf D55} (1997)5112,
    hep-th/9610043.

\bm{ewv} C. Vafa and E. Witten, {\it One-loop test of string duality},
    Nucl. Phys. {\bf B447} (1995) 261, hep-th/9505053.

\bm{dlm} M.J. Duff, J.T. Liu and R. Minasian, {\it Eleven dimensional origin
        of string-string duality: a one-loop test}, Nucl. Phys. {\bf 452}
    (1995) 261, hep-th/9506126.

\bm{mg} M.B. Green, {\it Connections between M-theory and superstrings},
    hep-th/9712195.

\bm{psh} P.S. Howe, {\it Weyl superspace},  Phys. Lett. {\bf B415} (1997) 149,
    hep-th/9707184.

\bm{bn} B.E.W. Nilsson, {\it A superspace approach to branes and supergravity}, in
    {\it Theory of Elementary Particles}, eds. H. Dorn, D. L\"ust and
    G. Weight (Wiley-Vch, 1998).


\bm{pktint} G. Papadopoulos and P.K. Townsend, {\it Intersecting M-branes},
    Phys. Lett. {\bf B380} (1996) 273, hep-th/9603087.

\bm{berint} E. Bergshoeff, M. de Roo, E. Eyras, B. Janssen and J. P. van der
    Schaar, {\it Multiple intersections of D-branes and M-branes},
    Nucl.Phys. B494 (1997) 119, hep-th/9612095.

\bm{gaunt} J.P. Gauntlett, {\it Intersecting branes}, hep-th/9705011.

\bm{hull} C.M. Hull, {\it Gravitational duality, branes and charges},
    Nucl. Phys. {\bf B509} (1998) 216, hep-th/9705162.

\bm{ggp} G.W. Gibbons, M.B. Green and M.J. Perry, {\it Instantons and
    seven-branes in type IIB superstring theory}, Phys. Lett. {\bf B370}
    (1996) 37, hep-th/9511080.

\bm{mo} P. Meessen and T. Ortin, {\it An Sl(2,Z) multiplet of nine-dimensional
        type II supergravity theories}, hep-th/9806120.

\bm{eyras2} E. Eyras, B. Janssen and Y. Lozano, {\it  5-branes, KK-monopoles
    and T-duality}, hep-th/9806169.

\bm{ber2} E. Bergshoeff, B. Janssen and T. Ortin, {\it Kaluza-Klein
    monopoles and gauged sigma models}, Phys. Lett. {\bf B410} (1997) 131,
    hep-th/9706117.

\bm{ber3} E. Bergshoeff, J.P. van der Schaar, {\it On M9 branes},
    hep-th/9806069.

\bm{mandm} A. Losev, G. Moore and S.L. Shatashvili, {\it M \& m's},
    Nucl. Phys. {\bf B522} (1998) 105, hep-th/9707250.

\bm{ewflux} E. Witten, {\it On flux quantization in M-theory and the effective
    action}, J. Geom. Phys. {\bf 22} (1997) 1, hep-th/9609122.


\bm{e1} D. Sorokin, V. Tkach and D.V. Volkov , {\it Superparticles,
    twistors and Siegel symmetry},Mod. Phys. Lett. {\bf A4} (1989) 901.

\bm{e2} D. Sorokin, V. Tkach, D.V. Volkov and A. Zheltukhin, {\it From
    superparticle Siegel supersymmetry to the spinning particle proper-time
    supersymmetry}, Phys. Lett. {\bf B216} (1989) 302.

\bm{bsm2} I. Bandos, P. Pasti, D. Sorokin, M. Tonin and D. Volkov, {\it
    Superstrings and supermembranes in the doubly supersymmetric geometrical
    approach}, Nucl. Phys. B446 (1995) 79, hep-th/9501113.

\bm{a1} I. Bandos, D.Sorokin and D. Volkov, {\it On the generalized
    action principle for superstrings and supermembranes},
    Phys. Lett. {\bf B352} (1995) 269, hep-th/9502141.

\bm{hs1} P.S. Howe and E. Sezgin, {\it Superbranes}, Phys. Lett. {\bf
        B390} (1997) 133, hep-th/9607227.

\bm{hs2} P.S. Howe and E. Sezgin, {\it D=11,p=5}, Phys. Lett. {\bf B394}
    (1997) 62, hep-th/9611008.

\bm{hsw1} P.S. Howe, E. Sezgin and P.C. West, {\it Covariant field
    equations of the $M$-theory five-brane}, Phys. Lett. {\bf 399B} (1997)
    49, hep-th/9702008.

\bm{hsw3} P.S. Howe, E. Sezgin and P.C. West, {\it Aspects of
    superembeddings}, hep-th/9705093.

\bm{cs} C.S. Chu and E. Sezgin, {\it M-Fivebrane from the open
    supermembrane}, JHEP {\bf 12} (1997) 001, hep-th/9710223.

\bm{chs} C.S. Chu, P.S. Howe and E. Sezgin, {\it Strings and D-branes
    with boundaries}, hep-th/9801202.

\bm{chsw}  C.S. Chu, P.S. Howe, E. Sezgin and P.C. West, {\it Open
    superbranes}, hep-th/9803041.

\bm{hrs2} P.S. Howe, O. Raetzel and E. Sezgin, {\it On brane actions
    and superembeddings}, hep-th/9804051.

\bm{db} P.S. Howe, O. Raetzel, E. Sezgin and P. Sundell, {\it D-brane
    superembeddings}, in preparation.

\bm{lb} P.S. Howe, O. Raetzel, I. Rudychev and E. Sezgin,
    {\it L-branes}, in preparation.

\bm{b1} T. Adawi, M. Cederwall, U. Gran, M. Holm and B.E.W. Nilsson, {\it
    Superembeddings, nonlinear supersymmetry and 5-branes}, hep-th/9711203.


\bm{dirachs} P.A.M. Dirac, {\it A remarkable representation of the 3+2 de
    Sitter group}, J. Math. Phys. {\bf 4} (1963) 901.

\bm{ff1} M. Flato and C. Fronsdal, {\it One massless particle equals two
    Dirac singletons}, Lett. Math. Phys. {\bf 2} (1978) 421.

\bm{ff3} M. Flato and C. Fronsdal, {\it Quantum field theory of
    singletons. The Rac}, J. Math. Phys. {\bf 22} (1981) 1100.

\bm{ns} H. Nicolai and E. Sezgin, {\it Singleton representations of
    $OSp(N|4)$}, \pl{143}{84}389.

\bm{es1} E. Sezgin, {\it The spectrum of the eleven dimensional
    supergravity compactified on the round seven sphere}, Trieste preprint
    IC-83-220, Nov 1983, in Phys. Lett. {\bf B138} (1984) 57, and
    in {\it Supergravities in Diverse Dimensions}, eds. A. Salam and E.
    Sezgin (World Scientific, 1988).

\bm{g1} M. G\"{u}naydin, {\it Oscillator-like unitary representations of
    noncompact groups and supergroups and extended supergravity theories},
    in {\it Lecture Notes in Physics, Vol. 180} (1983), eds.
    E. In\"on\"u and M. Serdaro\'glu.

\bm{es2} E. Sezgin, {\it The spectrum of D=11 supergravity via harmonic
    expansions on $S^4 \times S^7$}, Fortsch. Phys. {\bf 34} (1986) 217.

\bm{g6} M. G\"{u}naydin, L.J. Romans and N.P. Warner, {\it Spectrum
    generating algebras in Kaluza-Klein theories}, Phys. Lett. {\bf B146}
    (1984) 401.

\bm{duff1} M.J. Duff, {\it Supermembranes: the first fifteen weeks},
    Class. Quantum Grav. {\bf 5} (1988) 189.

\bm{bd1} M.P. Blencowe and M.J. Duff, {\it Supersingletons},
     Phys. Lett. {\bf B203} (1988) 229.

\bm{nst} H. Nicolai, E. Sezgin and Y. Tanii, {\it Conformally invariant
    supersymmetric field theories on $S^p\times S^1$ and super $p$-branes},
    \np{305}{88}483.

\bm{bst00} E. Bergshoeff, E. Sezgin and Y. Tanii, {\it A quantum
    consistent supermembrane theory}, Trieste preprint, IC/88/5. Jan
    1988.

\bm{mc1} E. Bergshoeff, M.J. Duff, C.N. Pope and E. Sezgin {\it
    Supersymmetric supermembrane vacua and singletons}, \pl{199}{87}69.

\bm{mc2} E. Bergshoeff, M.J. Duff, C.N. Pope and E. Sezgin, {\it
         Compactifications of the eleven-dimensional supermembrane},
         Phys. Lett. {\bf B224} (1989) 71.

\bm{dps} M.J. Duff, C.N. Pope and E. Sezgin, {\it A stable supermembrane
    vacuum with a discrete spectrum}, \pl{225}{89}319.

\bm{ope} E. Bergshoeff, E. Sezgin and Y. Tanii, {\it Stress tensor commutators
    and Schwinger terms in singleton theories}, Int. J. Mod. Phys.
    {\bf A5} (1990)3599.

\bm{st} E. Sezgin and Y. Tanii, {\it Superconformal sigma models in higher
    than two dimensions}, Nucl. Phys. {\bf B443} (1995) 70, hep-th/9412163.

\bm{phu} P.S. Howe, unpublished.

\bm {rk2} P. Claus, R. Kallosh, J. Kumar, P.K. Townsend and A. Van Proeyen, {\it
    Conformal theory of M2, D3, M5 and `D1+D5' branes}, hep-th/9801206.

\bm{rk3} R. Kallosh, J. Rahmfeld and A. Rajaraman, {\it Near horizon
    superspace}, hep-th/9805217.

\bm{fre} G. Dall'Agata, D. Fabbri, C. Fraser, P. Fr\'e, P. Termonia and M.
    Trigiante,{\it The $Osp(8|4)$ singleton action from the supermembrane},
    hep-th/9807115.

\bm{dewit} B. de Wit, K. Peeters, J. Plefka and A. Sevrin {\it The M-theory
    two-brane in $AdS_4 \times S^7$ and $AdS_7 \times S^4$},
        hep-th/9808052.

\bm{fv0} E.S. Fradkin and M.A. Vasiliev, {\it On the gravitational
    interaction of massless higher spin fields}, Phys. Lett. {\bf B189}
    (1987) 89.

\bm{fv1} E.S. Fradkin and M.A. Vasiliev, {\it Cubic interaction in
    extended theories of massless higher-spin fields}, Nucl. Phys.
    {\bf B291} (1987) 141.

\bm{v4} M.A. Vasiliev, {\it Consistent equations for interacting
    massless fields of all spins in the first order curvatures}, Ann. Phys.
    {\bf 190} (1989) 59.

\bm{v10} M.A. Vasiliev, {\it More on equations of motion for interacting
    massless fields of all spins in $3+1$ dimensions}, Phys. Lett.
    {\bf B285} (1992) 225.

\bm{v11} M.A. Vasiliev, {\it Higher-spin gauge theories in four, three
    and two dimensions}, hep-th/9611024.

\bm{pers} E. Sezgin and P. Sundell, {\it Higher spin $N=8$ supergravity},
      to appear in Nuclear Phys. B, hep-th/9805125.

\bm{dn2} B. de Wit and H. Nicolai, {\it $N=8$ supergravity},
    Nucl. Phys. {\bf B208} (1982) 323.

\bm{cj2} E. Cremmer and B. Julia, {\it The SO(8) supergravity},
     Nucl. Phys. {\bf B159} (1979) 141.

\bm{kv1} S.E. Konstein and M.A. Vasiliev, {\it Massless representations
    and admissibility condition for higher spin superalgebras}, Nucl. Phys.
    {\bf B312} (1989) 402.

\bm{vhvp} J.W. van Holten and A. van Proeyen, {\it N=1 supersymmetry
      algebras in d=2,3,4 mod 8}, J. Phys. {\bf A15} (1982) 763.

\bm{g5} M. G\"{u}naydin, {\it Unitary supermultiplets of OSp(1/32,R) and
    M-theory},  hep-th/9803138.

\bm{df1} R. D'Auria and P. Fr\'e, {\it Geometric supergravity in D=11
    and its hidden supergroup}, Nucl. Phys. {\bf B201} (1982) 101.

\bm{df2} L. Castellani, P. Fr\'e, F. Giani, K. Pilch and P. van Nieuwenhuizen,
     {\it Gauging of d=11 supergravity?}, Ann. Phys. {\bf 146} (1983) 35.

\bm{malda} J.M. Maldacena, {\it The large N limit of superconformal field
    theories and supergravity}, hep-th/9711200.

\bm{gkp} S.S. Gubser, I.R. Klebanov and A.M. Polyakov, {\it Gauge theory
    correlators from non-critical string theory}, hep-th/9802109.

\bm{ew} E. Witten, {\it Anti-de Sitter space and holography},
        hep-th/9802150.


\bm{dnp} M.J. Duff, B.E.W. Nilsson and C.N. Pope, {\it Kaluza-Klein
    supergravity}, Phys. Rep. {\bf 130}(1986)1.

\bm{conf} M.J. Duff, C.N. Pope and E. Sezgin (eds) {\it Supermembranes and
    Physics in (2+1)-Dimensions}(World Scientific, 1990), Proceedings of
    Trieste Conference, Trieste, July 1989.

\bm{caracas} E. Sezgin, {\it Comments on supermembrane theories}, in {\it
    Strings, Membranes and QED}, eds. C. Aragone, A. Restuccia and S.
    Salam\'o (Universidad Sim\'on Bol\'ivar, 1992).

\bm{giveon} A. Giveon, M. Porrati and E. Rabinovici, {\it Target space duality
    in string theory}, Phys. Rep. {\bf 244} (1994) 77, hep-th/9401139.

\bm{dkl}M.J. Duff, R.R. Khuri and J.X. Lu, {\it String solitons}, Phys. Rep.
    {\bf 259} (1995) 213, hep-th/9412184.

\bm{js2} J.H. Schwarz, {\it Lectures on superstrings and M theory dualities},
    hep-th/9607201.

\bm{pol2} J. Polchinski, {\it TASI lectures on D-branes}, hep-th/9611050.

\bm{duff96} M.J. Duff, {\it Supermembranes}, hep-th/9611203.

\bm{pktr} P.K. Townsend, {\it Four lectures on M-theory}, hep-th/9612121.

\bm{sen} A. Sen, {\it An introduction to non-perturbative string theory},
    hep-th/9802051

\bm{duffrev} M.J. Duff, {\it  A layman's guide to M theory}, hep-th/9805177.

\bm{jhs3} J.H. Schwarz, {\it From superstrings to M theory}, hep-th/9807135.

\bm{duffads} M.J. Duff, {\it Ani-de Sitter space, branes, singletons,
    superconformal field theories and all that}, hep-th/9808100.

\fin